\documentclass[11pt]{article}

\usepackage{latexsym,times}
\usepackage{amsmath,amsfonts,mathrsfs}
\usepackage[pdftex]{hyperref}
\usepackage[inline]{enumitem}
\usepackage{xcolor}
\usepackage{graphicx}
\usepackage{framed}
\usepackage[square,sort&compress,comma,numbers]{natbib}
\usepackage{subcaption}

\newcommand{\req}[1]	{(\ref{#1})}

\newcommand{\p}	{\partial}

\newcommand{\R}	{\mathbb{R}}

\newcommand{\N}	{\mathbb{N}}

\newcommand{\cA}{\mathcal{A}}
\newcommand{\cD}{\mathcal{D}}
\newcommand{\scD}{\mathscr{D}}
\newcommand{\cF}{\mathcal{F}}
\newcommand{\cG}{\mathcal{G}}
\newcommand{\cH}{\mathcal{H}}
\newcommand{\scH}{\mathscr{H}}

\newcommand{\cJ}{\mathcal{J}}
\newcommand{\cK}{\mathcal{K}}
\newcommand{\cM}{\mathcal{M}}
\newcommand{\cN}{\mathcal{N}}
\newcommand{\cO}{\mathcal{O}}

\newcommand{\cW}{\mathcal{W}}

\newcommand\sphere[1]   {\mathcal{A}_{#1}}

\newcommand{\be}	{\begin{equation}}
\newcommand{\ee}	{\end{equation}}
\newcommand{\ba}	{\begin{align}}
\newcommand{\ea}	{\end{align}}

\newcommand{\tr}	{\mathrm{tr}}

\newcommand\Algebra[1]	{\mathfrak{#1}}
\newcommand\Group[1]	{\mathrm{#1}}
\newcommand{\SO}	{\Group{SO}}

\newcommand{\su}	{\Algebra{su}}
\newcommand{\hs}	{\operatorname{hs}}

\newcommand\ket[1]{\vert{#1}\rangle}

\newcommand{\corr}[1]{\left\langle{#1}\right\rangle}

\newcommand\cft {\mathsf{CFT}}
\newcommand\ads {\mathsf{AdS}}
\newcommand\vol {\operatorname{vol}}

\newcommand\hg[4]{{}_2F_1\biggl(\!\!\begin{array}{c}{#1},{#2}\\{#3}\end{array}\!\!\Big|\,{#4}\biggr)}
\newcommand\htt[6]{{}_3F_2\biggl(\!\!\begin{array}{c}{#1},{#2},{#3}\\{#4},{#5}\end{array}\!\!\Big|\,{#6}\biggr)}
\newcommand\ghg[5]{{}_{#1}F_{#2}\biggl(\!\!\begin{array}{c}{#3}\\{#4}\end{array}\!\!\Big|\,{#5}\biggr)}
\newcommand\f[2]{{}_{3}f_{2}\biggl(\!\!\begin{array}{c}{#1}\\{#2}\end{array}\!\!\biggr)}

\newcommand\half{{\tfrac{1}{2}}}

\newcommand{\ishiket}[1]{|#1\rangle\!\rangle}

\newcommand{\braneket}[1]{\|#1\rangle\!\rangle}

\newcommand{\cor}[1]{\left\langle #1 \right\rangle}

\newcommand\ppu     {u} 
\newcommand\hyp     {H}

\makeatletter

\@addtoreset{equation}{section} 
\makeatother

\begin{document}
\begin{titlepage}
\title{Double Trace Interfaces}
\author{
{\sc Charles M. Melby-Thompson${}^a$} and
{\sc Cornelius Schmidt-Colinet${}^b$}}
\date{}
\maketitle \vspace{-1.0cm}
\begin{center}
~\\
{\it 
${}^{a}$
\parbox[t]{0.6\textwidth}{Department of Physics, Fudan University\\
220 Handan Road, 200433 Shanghai, China}
} \\
\medskip
{\it
${}^{b}$
\parbox[t]{0.6\textwidth}{Arnold Sommerfeld Center, Ludwig-Maximilians-Universit\"at\\
\quad Theresienstra{\ss}e 37, 80333 M\"unchen, Germany}
}
\\
\bigskip
charlesmelby@gmail.com
\qquad\quad
cornelius.sc@gmail.com
\end{center}

\begin{abstract}
\vskip0.1cm  \noindent 
We introduce and study renormalization group interfaces between two holographic conformal theories which are related by deformation by a scalar double trace operator.
At leading order in the $1/N$ expansion, we derive expressions for the two point correlation functions of the scalar, as well as the spectrum of operators living on the interface.
We also compute the 
interface contribution to the sphere partition function,
which in two dimensions gives the boundary $g$ factor.
Checks of our proposal include reproducing the $g$ factor and some defect overlap coefficients of Gaiotto's RG interfaces at large $N$, and the two-point correlation function whenever conformal perturbation theory is valid.
\end{abstract} 

{\small
\begin{flushleft}
~~\\
\end{flushleft}}
\thispagestyle{empty}
\end{titlepage}
\newpage
\tableofcontents

\section{Introduction}
\label{sec:introduction}
Conformal defects have played an important role in the development of conformal field theory (CFT).
Of particular interest for many purposes are conformal interfaces, those interfaces separating two different CFTs that preserve a maximal subgroup of the conformal group.

A particularly interesting class of interfaces is given by renormalization group (RG) interfaces~\cite{Brunner:2007ur}, which are associated to a renormalization group flow from $\cft_1$ to $\cft_2$. 
In addition to being of intrinsic interest, such defects may provide new tools to study the behavior of renormalization group flows.
Various such interfaces, both approximate~\cite{Konechny:2014opa,Brunner:2015vva,Gliozzi:2015qsa} and
(in the presence of supersymmetry) exact~\cite{Brunner:2007ur,Gaiotto:2012np,Dimofte:2013lba} have been 
constructed, but in general it is difficult to compute observables that are not protected by symmetry.
In particular, we are not aware of computations of two-point correlation functions in the particular case of RG interfaces.

Within the AdS/CFT correspondence, conformal interfaces are typically realized using the Janus construction~\cite{Bak:2003jk},%
\footnote{Bulk D-branes can also play the role of interfaces. This can be understood as the thin wall limit of the Janus construction.}
for which various approximate and (in the supersymmetric case) exact solutions are known (see \textit{e.g.}~\cite{DHoker:2008lup,DHoker:2009lky,Nishioka:2010ha,Bak:2011ga})''. 
The construction takes advantage of the $\SO(d,1)$ symmetry preserved by the interface to slice the bulk geometry by copies of hyperbolic space,%
\footnote{In this paper we work exclusively in Euclidean signature, in which case the vacuum bulk geometry is $(d+1)$-dimensional hyperbolic space $H^{d+1}$, and the conformal group is $SO(d+1,1)$.}
\be
ds^2 = d\beta^2 + f(\beta) ds_{H^d}^2 \,.
\ee
Pure hyperbolic space corresponds to $f(\beta)=\cosh^2\!\beta$.
The deformation of $f(y)$ away from this is sourced by scalar field gradients $\phi(\beta)$, the details of which depend on the scalar potential.
The bulk equations of motion are in general difficult to solve, and even in those cases where solutions are available, simple observables such as two-point correlation functions are difficult to compute: the only computation of a (non-protected) holographic two-point function we are aware of was performed in~\cite{Chiodaroli:2016jod}. 

Holographic realizations of RG interfaces have appeared in the literature: see, for example, \cite{Bobev:2013yra,Karndumri:2017bqi}.
The purpose of this paper is to introduce and study a type of holographic RG interface that we refer to as (holographic) \textit{double trace interfaces}. 
It has been known since the work of \cite{Klebanov:1999tb} that whenever
the gravitational dual of a CFT has a scalar field whose mass lies in the 
unitarity window $-\frac{d^2}{4} \le m^2 \le -\frac{d^2}{4}+1$, there are
two consistent choices of boundary asymptotics.%
\footnote{It is possible and interesting to relax this assumption, but doing so breaks unitarity. We restrict ourselves in this paper to unitary theories.}
The two different choices lead to two different CFTs on the boundary, with different spectra.
For one choice, the scalar field $\phi$ is dual to a gauge-invariant (single trace) operator $\varphi_+$
of dimension $\Delta_+$, while the other choice leads to an operator 
$\varphi_-$ of dimension $\Delta_-$.
These two CFTs are related by RG flow from $\cft_-$ to $\cft_+$, which is initiated on the CFT side through perturbation by the ``double trace'' operator $(\varphi_-)^2$.
This is implemented holographically by imposing on the scalar field boundary conditions of mixed Dirichlet-Neumann type; 
renormalization group flow from the UV to the IR is realized in terms of the dominant asymptotics in the near-boundary and deep bulk regions, respectively. 
These RG flows are particularly simple at large $N$.
This can be traced to the fact that their effects are due entirely to the asymptotics of quantum fluctuations, and as a result, gravitational backreaction occurs only at loop level.
As a result, the leading contribution to any computation takes place on a pure AdS
background. 
This fact makes feasible, at least at leading order, computations that are impractical in the general case.

\begin{figure}
\centering
\begin{subfigure}[b]{0.5\textwidth}
    \centering
    \includegraphics[width=0.5\linewidth]{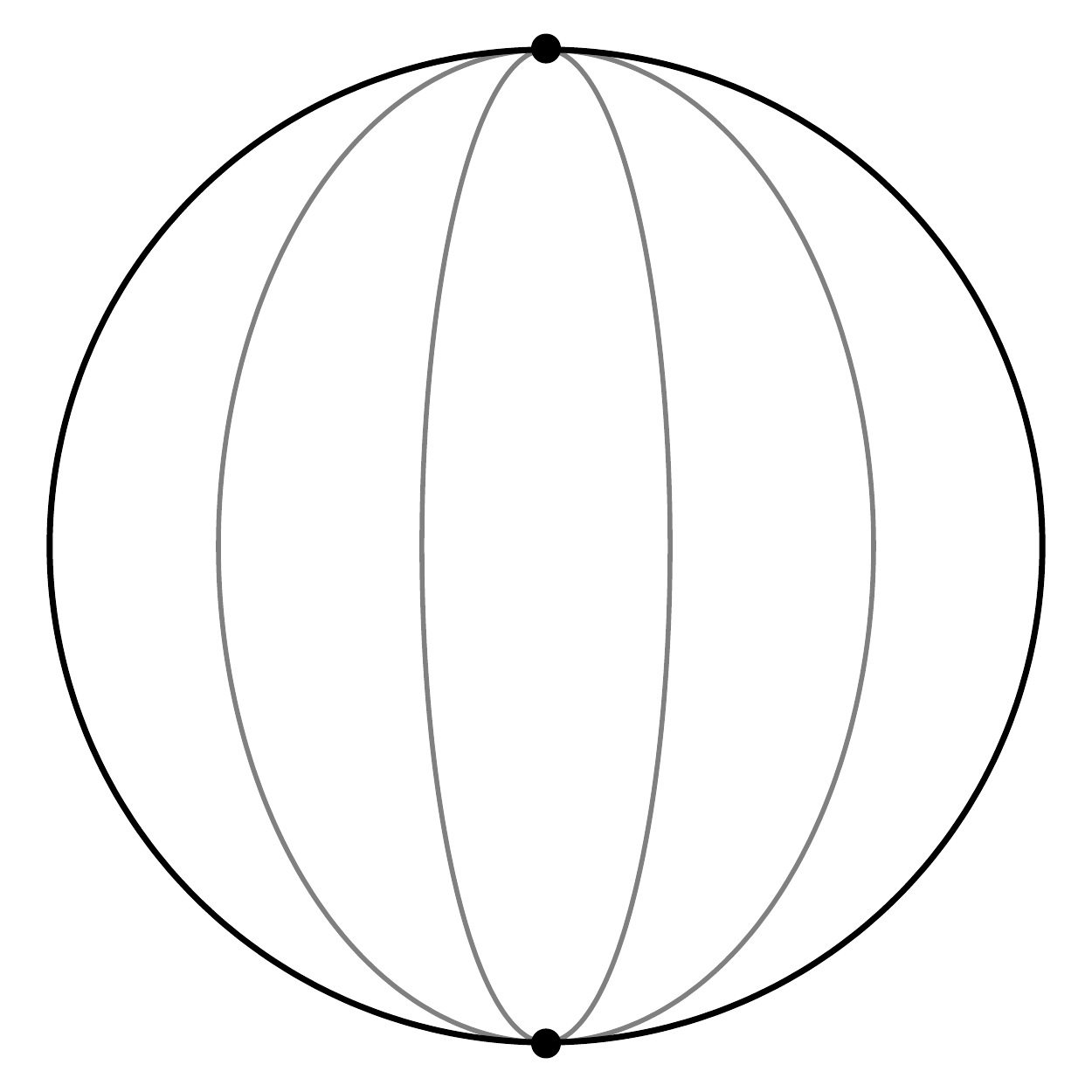}
    \caption{}
    \label{fig:janus}
\end{subfigure}
\begin{subfigure}[b]{0.48\textwidth}
    \centering
    \begin{picture}(108,108)(0,0)
        \put(0,0){\includegraphics[width=.5\linewidth]{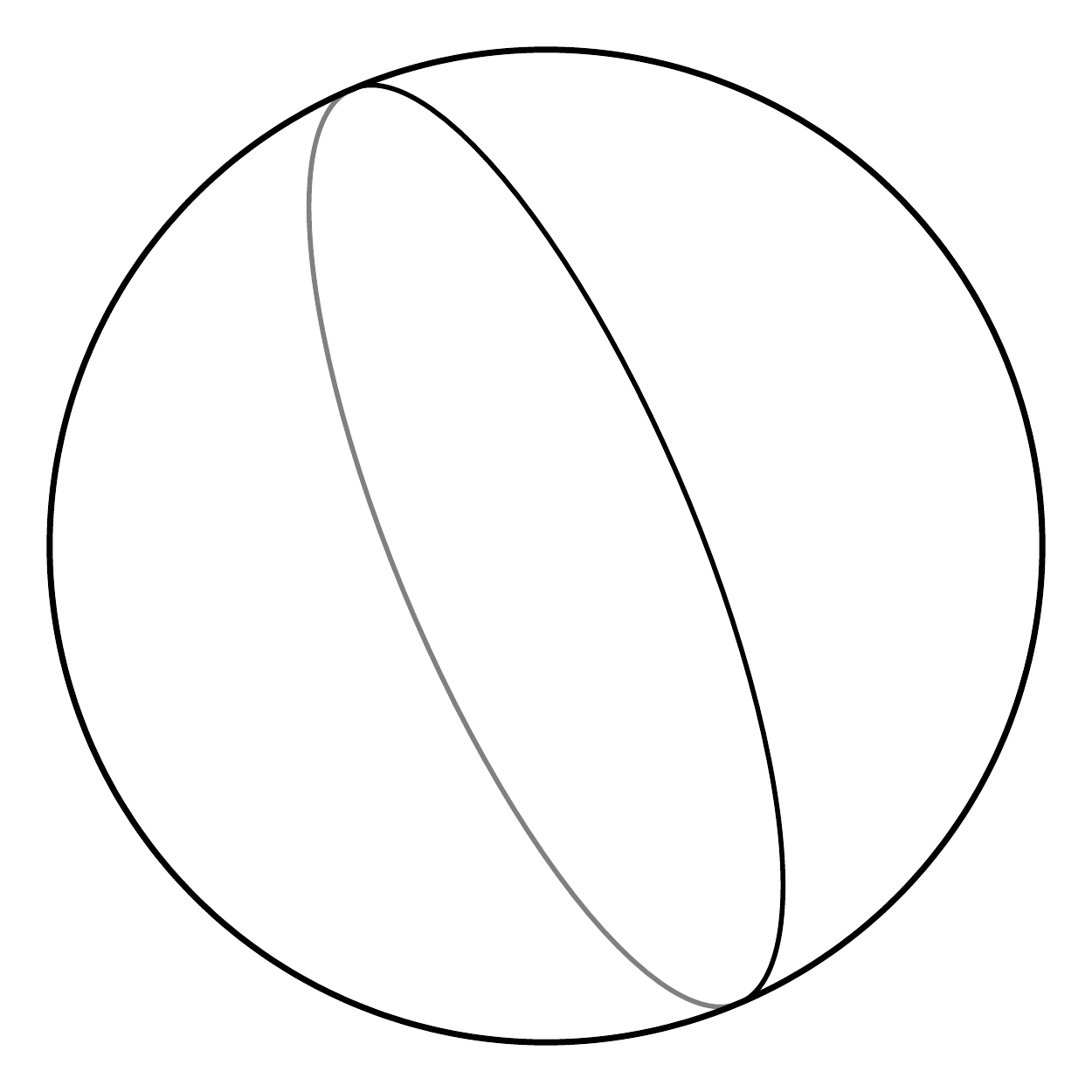}}
        \put(7,20){\makebox(0,0){$u^{\Delta_-}$}}
        \put(105,102){\makebox(0,0){$u^{\Delta_+}$}}
    \end{picture}
    \caption{}
    \label{fig:mixed}
\end{subfigure}%
\caption{(a) Janus coordinates: $H^{d+1}$ is sliced by copies of $H^d$, which intersect along the defect. 
(b) Mixed boundary conditions on the two halves of the boundary of $H^{d+1}$.
\label{defect-figure}}
\end{figure}

Consider hyperbolic space $H^{d+1}$, with its boundary divided into two regions, $A_+$ and $A_-$
(figure~\ref{defect-figure}), where the local physics is described by $\cft_+$ and $\cft_-$ respectively. 
Near boundary region $A_+$, the quantum fluctuations of the bulk scalar field $\phi$ should have scaling dimension $\Delta_+$, while those near boundary region $A_-$ should have dimension $\Delta_-$.
The asymptotics of quantum fluctuations contribute to diagrammatic computations through the particular choice of bulk Green's function $G$ for $\phi$. 
The choice of Green's function is therefore what determines all properties of the holographic interface, and sections~\ref{sec:dti} and \ref{sec:integral transform} are devoted to its analysis.

With the Green's function in hand, in principle all observables associated to the interfaces can be computed by using this Green's function in all Witten diagrams. 
In this paper, we focus on the simplest observables that can be derived from $G$: (1) the two-point correlation function at tree level, and (2) the one-loop partition function. 
From the correlation function one can further extract the spectrum of non-trivial defect operators.
The two-point function can also be compared with CFT results. 
In particular, we show that our bulk expressions reproduce results that we derive in conformal perturbation theory. 
Furthermore, from the conformal block expansion we can read off relations between the bulk and bulk-boundary OPE coefficients, which allows us to derive an expression for the defect overlap coefficients for certain operators in the large $N$ limit.
As an example, we reproduce the large $N$ behavior of overlap coefficients for the interfaces of~\cite{Gaiotto:2012np} between adjacent $\cW_N$ minimal models.
We further compute the contribution of Gaiotto's interface to the sphere partition function in the large $N$ limit, and show that it exactly matches the interface contribution to the bulk one-loop partition function.

\medskip
The logic of this paper is as follows.
Section~\ref{sec:dti} gives a detailed definition of holographic double trace interfaces, and discusses methods available for deriving the bulk Green's functions for an arbitrary interface geometry.
Section~\ref{sec:integral transform} turns to the explicit evaluation of the Green's function in the case of a spherical interface, which is done using the more powerful tools of harmonic analysis on $H^d$;
these tools also allow us to derive the spectrum of interface operators.
Section~\ref{sec:correlation functions} treats the evaluation of the CFT two-point function at leading order in the $1/N$ expansion, from which we extract the dimensions of a sequence of primary operators living on the interface, matching the results from section~\ref{sec:integral transform}.
Section~\ref{sec:boundary entropy} computes the leading contribution of the interface to the partition function by evaluating the one-loop vacuum bubble diagram.

Section~\ref{sec:cft} is devoted to computations in CFT, which provide two tests of our results. 
As the first, we derive the CFT two-point function in the presence of double trace interfaces within conformal perturbation theory, and show that it matches our bulk computation in parameter regimes where both descriptions are valid.
The second is to derive within the higher spin gravity/WCFT
duality of \cite{Gaberdiel:2010pz} the boundary $g$-factor and several overlap coefficients for the RG interfaces of~\cite{Gaiotto:2012np} joining the $\cW_{N,k}$ and $\cW_{N,k-1}$ minimal models.
We find that both match the results of sections~\ref{sec:correlation functions} and \ref{sec:boundary entropy}. 
We close with a summary of our conclusions and a list of interesting questions and problems for the future.

\section{Double trace interfaces%
\label{sec:dti}}
The construction of a double trace interface begins with a pair of $d$-dimensional unitary $\cft$s, $\cft_\pm$, which have dual descriptions in terms of a single gravitational theory on a weakly curved $\ads$ space, and are related by the choice of boundary condition for a bulk scalar field $\phi$ with $m^2 = -\frac{d^2}{4}+\nu$.
We take the mass to lie in the unitarity window, defined by $0< \nu<1$.
The two CFTs therefore differ at leading order in the $1/N$ expansion by the choice of dimension $\Delta_\pm=\frac{d}{2}\pm\nu$ for a single operator $\varphi_\pm$.

Our goal is to describe a conformal interface separating a region $A_+$ whose local physics is that of $\cft_+$, and the complementary region $A_-$ described by $\cft_-$.
How does one realize such an interface?
The AdS/CFT dictionary says that the field $\phi$ should have boundary condition $\Delta_+$ near the $\cft_+$ boundary, and boundary condition $\Delta_-$ near the $\cft_-$ region.
To be more precise about what we mean, consider Poincar\'e patch coordinates $X=(u,\chi)$ with metric%
\footnote{We work throughout in units such that the AdS length $\ell_{\ads}=1$.}
\be
ds^2_{H^{d+1}} = \frac{du^2 + ds^2_{\R^{d}}(\chi)}{u^2}
\,.
\ee
By boundary condition, we mean that any configuration of the field $\phi$ appearing in the path integral must fall off near the boundary as
\be
    \phi(u,\chi) = \left\{ 
        \begin{array}{ll}
            \Psi_+(\chi)u^{\Delta_+} + O(u^{\Delta_+ + 2})\,, \qquad & \chi\in A_+ \\
            \Psi_-(\chi)u^{\Delta_-} + O(u^{\Delta_- + 2})\,, & \chi\in A_- \,.
        \end{array}
    \right.
\ee
This is accomplished in Witten diagrams by making a particular choice of inverse for the kinetic operator.
The interface is therefore implemented in the bulk by choosing the appropriate bulk Green's function. 

To be explicit, a double trace interface is obtained by imposing the following conditions on the Green's function $G$:
\begin{enumerate}[label=(\Alph*)]
\item $G$ satisfies the defining equation%
\footnote{We use $\delta(x,y)$ to denote the covariant delta function,
$\delta(x,y)=\frac{1}{\sqrt g}\delta(x-y)$.}
\be
(-\Box+m^2)G(X,X') = \delta^{(d+1)}(X,X') \,.
\ee
\item As $X'=(u',\chi')$ approaches a boundary point $\chi'\in A_\pm$,
\be
    G(X;X') = \pm \frac{1}{2\nu} u'^{\Delta_\pm}K^\pm(X;\chi') + O(u'^{\Delta_\pm+2})
    \qquad\text{as}\; u'\to 0 \,.
\ee
\end{enumerate}
The form of $K^\pm$ is not important for this definition, but it is in fact the bulk-boundary propagator associated to the region $A_\pm$. 
The factor of $\pm\frac{1}{2\nu}$ is the standard prefactor $\frac{1}{2\Delta_\pm-d}$.

We consider here two methods of solving these conditions.
The first is harmonic analysis: when the bulk geometry can be expressed as a warped product of a symmetric space over an interval, the decomposition in terms of Laplacian eigenfunctions reduces the above equations to an ODE.
This method works whenever the wave equation is separable in coordinates respecting the boundary geometry of the defect, as happens when the interface is spherical or planar.
Otherwise, one must use the more general methods developed to deal with \emph{mixed boundary value problems} for partial differential equations; for a thorough treatment of this subject, see for example~\cite{Sneddon}. 
As a simple example, we will outline at the end of this section the application of such methods to the derivation of the bulk-boundary propagator in the case of spherical defects;
a full derivation using these methods is offered in appendix~\ref{sec:K as mbvp}.

\subsection{Double trace interfaces as a mixed boundary value problem}
The Green's function solves a boundary value problem in which the boundary is split into two regions $A_+$ and $A_-$, such that the function in question has Dirichlet-like boundary conditions on $A_+$, but Neumann-like boundary conditions on $A_-$.
Such problems are known as \emph{mixed boundary value problems}.
(This is not to be confused with ``mixed boundary conditions'', 
otherwise known as Robin boundary conditions, which refer to a 
spatially homogeneous linear combination of Dirichlet and Neumann 
boundary conditions.)

We begin by writing the mixed Green's function in the form
\be
G(X,X') = G_{\Delta_-}(X,X') + \cH(X,X') 
\label{GM-H}
\ee
where $G_{\Delta_-}$ is the homogeneous Green's function for $\Delta_-$ asymptotics.
Then $\cH$ satisfies the free scalar equation, so it can be written
as the convolution of a function on the boundary of $H^{d+1}$ with 
$K_{\Delta_-}(X,x')$, the bulk-boundary propagator for $\cft_-$.

Let $K^+(u,\chi;\chi')$ be the mixed bulk-boundary propagator associated to a boundary point $\chi'\in A_+$.
This function is determined by the following properties:
\begin{enumerate}[label={[}K\arabic*{]}]
\item \label{K condition 1}
    $(-\Box+m^2)K^+(u,\chi ; \chi')=0$,
\item \label{K condition 2}
    $[K^+]_{\Delta_-}(\chi ; \chi') = \delta(\chi,\chi')$ for $\chi\in A_+$,
\item \label{K condition 3}
    $[K^+]_{\Delta_+}(\chi ; \chi') = 0$ for $\chi\in A_-$.
\end{enumerate}
Here, by $[f]_{\Delta}$ we mean the coefficient of $u^\Delta$ in the expansion of $f$ as $u\to 0$.
$K^+$ is given in terms of the Green's function by the standard relation
\be
K^+(u,\chi;\chi')=\lim_{u'\to 0}\frac{2\nu}{u'^{\Delta_+}}G(u,\chi;u',\chi') \,,
\qquad
\chi'\in A_+ \,.
\ee
We claim that
\be
    \cH(X;X') 
    = \frac{1}{2\nu}\int_{A_+}d^d\chi''\, K^+(X;\chi'')K_{\Delta_-}(X';\chi'')
\label{eq:H from K}
\ee
where $K_{\Delta_-}$ is the bulk-boundary propagator for the $\Delta_-$ CFT.
Recalling that 
\be
[G_{\Delta_\pm}]_{\Delta_\pm} = \pm \frac{1}{2\nu}K_{\Delta_\pm}
\ee
it is straightforward to verify that as a function of $(u,\chi)$,
$H$ satisfies the asymptotic conditions
\begin{align}
[\cH]_{\Delta_-}(\chi;u',\chi') &= -[G_{\Delta_-}]_{\Delta_-} && \chi\in A_+ 
\\
[\cH]_{\Delta_+}(\chi;u',\chi') &= 0 && \chi\in A_- \,.
\end{align}
Defining $G$ as in \req{GM-H} implies that it satisfies both conditions (A) and (B).
Equations \req{GM-H} and \req{eq:H from K} therefore express $G$ in terms of $K^+$, reducing the problem to solving \ref{K condition 1}-\ref{K condition 3}.

\bigskip
We see thus that observables of double trace interfaces can be expressed in terms of the bulk-boundary propagator, and thus it is this object that will be the primary focus of what follows.
We focus in particular on the case of a spherical defect.
This case is special because it preserves a maximal subgroup of the conformal group, allowing us to solve the problem as an ODE using harmonic analysis on $H^d$.
This is done in section~\ref{sec:integral transform}.
For any other shape, it is necessary to solve for $K$ as a mixed boundary value problem.
To illustrate this process, we show in detail how this can be done for the spherical interface in appendix~\ref{sec:K as mbvp}.

The remainder of the section will be occupied with holographic renormalization and the extraction of correlation functions in section~\ref{sec:corr}, and some comments on the case of general interface shapes in section~\ref{sec:defect fusion}.

\subsection{Holographic renormalization and correlation functions}
\label{sec:corr}
Let us now consider the question of how to extract correlation functions from the bulk-boundary propagator associated to a general double trace interface. 
The AdS/CFT dictionary states that for each bulk field $\phi$ dual to a scalar operator $\varphi$, the solution to the equations of motion can be expanded in the form
\be
\phi(u,\chi) = a_J J(\chi) u^{\Delta_J} + \psi(\chi)u^{\Delta_\varphi} + \cdots
\label{eq:phi expansion}
\ee
where $J$ denotes a source, $\psi(\chi)$ is proportional to the one-point function $\corr{\varphi(\chi)}_J$ in the presence of $J$, and all other terms are local functionals of $J$ and $\varphi$;
$a_J$ is a free parameter that we will fix later.
Note $\Delta_J + \Delta_{\varphi} = d$.
$J$ and $\psi$ are locally independent, but are determined by each other upon requiring non-singular behavior in the bulk.
Correlation functions are obtained by the statement that the gravitational partition function with boundary conditions $J$ is equal to the generating functional of the CFT with source $J$:
\be
Z_\text{CFT}(J) = 
Z_\text{gravity}(\phi\sim J u^{\Delta_J}+\cdots) \,.
\ee
Defining $W=\log Z$, the connected correlation functions are
\be
\corr{\varphi(\chi)}_J 
= a_{\varphi} \psi(\chi)\vert_J 
= \frac{\delta W(J)}{\delta J(\chi)}\,
\qquad\quad
\corr{\varphi(\chi)\varphi(\chi')}_J 
= \frac{\delta W(J)}{\delta J(\chi)\delta J(\chi')} \,,
\label{eq:one-point}
\ee
where the value of $a_\varphi$ is determined by the effective action.
Note that, due to the presence of $a_J$ in~\req{eq:phi expansion}, this equation differs from the standard one by a factor of $a_J$.
This is a matter of the normalization of the operator dual to $J$.
It would be most natural to choose $a_J$ such that $a_\varphi=1$.
The standard normalization, however, sets $a_J=1$.
If $\phi$ has the $\Delta_-$ quantization, $a_\varphi$ is negative, which flips the sign of certain correlators relative to the natural expectation in CFT. 
Because the $a_\varphi=1$ normalization is ubiquitous in the literature, we choose $a_J=1$ for the $\Delta_+$ quantization; to obtain the natural sign for the mixed two-point functions, we therefore choose $a_J=-1$ for the $\Delta_-$ quantization.
We will see in section~\ref{sec:correlation functions} that this convention reproduces the sign of $\corr{\varphi_+\varphi_-}$ that is natural in conformal perturbation theory.

In the semi-classical limit $W$ is expressed in terms of the on-shell classical gravitational action $S_\text{os}$, $W=-S_\text{os}$, so this is the quantity we deal with for the rest of the section.
To render the variations well-defined, one requires a well-behaved variational principle.
In particular, this implies that if $\phi = \phi_c + \delta\phi$, where $\phi_c$ solves the bulk equations and $\delta\phi$ has $u^{\Delta_\varphi}$ asymptotics, then the variation of the action must be finite.
As is well known, to accomplish this requires the inclusion of local counterterms (holographic renormalization), and the counterterms we add determine the allowed fluctuations.

Since our system involves both boundary conditions for $\phi$, let us first briefly review how this works when there is no interface.
We restrict to $0<\nu<1$ as before, and expand near $u=0$ in the form
\be
\phi(u,\chi) = \phi_-(\chi) u^{\Delta_-} + \phi_+(\chi) u^{\Delta_+} + \cdots
\label{eq:field expansion}
\ee
where $(\cdots)$ is irrelevant to what follows.
Start with the variation of the bare on-shell action. 
Introduce a cutoff surface $u=\epsilon$, and let $S_\epsilon(\phi)$ be the cut-off bulk action.
As usual, for $\phi$ on-shell we write
\be
\delta S_\epsilon(\phi) = \int d^{d+1}\!X\sqrt{g}\bigl( \nabla\phi\cdot\nabla\delta\phi + m^2\Phi\delta\Phi \bigr)
= \int_{u=\epsilon}\!\!d^d\chi\sqrt{\gamma}\,\delta\phi\p_{\hat n}\phi
\,,
\ee
where $\gamma$ is the induced metric on the cutoff surface, $n$ is the outward-pointing unit normal, and we have dropped the term proportional to the equations of motion.
Expanding in $\epsilon$
(`$\simeq$' means up to terms that vanish as $\epsilon\to 0$), we find
\be
\delta S_\epsilon(\phi) \simeq \int d^d\chi\,
\Bigl( \Delta_- \phi_-\delta\phi_- \epsilon^{-2\nu} 
 + \Delta_+ \phi_+\delta\phi_- + \Delta_-\phi_-\delta\phi_+ \Bigr) \,.
\ee
We now add counterterms, which must render the variation finite.
Furthermore, if we want $\Delta_+$ boundary conditions, then the variation of the action should depend only on $\delta\phi_-$, while for $\Delta_-$ boundary conditions, it should depend on $\delta\phi_+$ only.
The first can be accomplished by the counterterm
\be
S^{\Delta_+}_\text{ct}(\phi) = \frac{\Delta_-}{2}\int_{u=\epsilon}\!\! d^d\chi\sqrt{\gamma}\phi^2
\quad\implies\quad
\delta S^{\Delta_+}_\epsilon(\phi) \simeq \int d^d\chi\bigl(
	\Delta_-\phi_-\delta\phi_-\epsilon^{-2\nu} + 2\Delta_-\phi_+\phi_-
\bigr)
\ee
which leads to 
\be
\delta S^{\Delta_+} = \lim_{\epsilon\to 0}(\delta S_\epsilon + \delta S_\text{ct}^{\Delta_+})
= \int d^d\chi (-2\nu)\phi_+\delta\phi_- \,.
\ee
Note that this gives $a_{\varphi}=-2\nu$.
We can obtain $\Delta_-$ boundary conditions by instead using the counterterm
\be
S^{\Delta_-}_\text{ct} = \frac{1}{\Delta_-}\int_{u=\epsilon} d^d\chi \sqrt{\gamma}(\p_n\phi)^2
\quad\implies\quad
\delta S^{\Delta_-}_\text{ct} \simeq \int d^d\chi\bigl(
	\Delta_- \phi_-\delta\phi_-\epsilon^{-2\nu}
	+ \Delta_+\delta(\phi_+\phi_-)
\bigr)
\ee
which gives
\be
\delta S^{\Delta_-} = \lim_{\epsilon\to 0}(\delta S_\epsilon + \delta S_\text{ct}^{\Delta_-})
= \int d^d\chi (2\nu)\phi_-\delta\phi_+ \,.
\ee

The bulk values of $\phi$ are determined by either one of $\phi_+$ or $\phi_-$ in terms of the bulk-boundary propagator:
\be
\phi(u,\chi) = \int d^d\chi'\, K_{\Delta_+}(u,\chi;\chi') \phi_-(\chi')
= -\int d^d\chi'\, K_{\Delta_-}(u,\chi;\chi')\phi_+(\chi')
\ee
(the minus sign is due to $a_{J_-}=-1$).
In $\cft_+$, the source is $J_+=\phi_-$, and $\phi_+$ is given by the relation
\be
\phi_+ = \int d^d\chi'\, [K_{\Delta_+}]_{\Delta_+}(\chi;\chi') 
\phi_-(\chi') \,.
\ee
From this we may obtain the standard result for the $\cft_+$ two-point function:
%
%
%
%
%
\begin{align}
\corr{\varphi_+(\chi)\varphi_+(\chi')}_{\cft_+} &= \frac{\delta^2(-S_\text{os})}{\delta\phi_-(\chi)\delta\phi_-(\chi')} = 2\nu\, [K_{\Delta_+}]_{\Delta_+}(\chi;\chi') \,.
\end{align}
The same applied to $\cft_-$ (with $J_-=-\phi_+$) gives the usual value
\begin{align}
\corr{\varphi_-(\chi)\varphi_-(\chi')}_{\cft_-} &= 
\frac{\delta^2(-S_\text{os})}{\delta\phi_+(\chi)\delta\phi_+(\chi')} = -2\nu\, [K_{\Delta_-}]_{\Delta_-} (\chi;\chi') \,.
\end{align}

\medskip
Let us now turn to our case of interest, where the boundary is divided into a region $A_+$ of $\cft_+$ and a region $A_-$ of $\cft_-$.
We can still expand any on-shell field configuration $\phi$ as in equation~\req{eq:field expansion}.
Our counterterms, and thus our identification of sources, is however different.
We must therefore use the counterterm $S^{\Delta_+}_\text{ct}$ in $A_+$, and $S^{\Delta_-}_\text{ct}$ in $A_-$.
Using this counterterm, the variation of the on-shell action becomes:
\be
\delta S_\text{os} = \int_{A_+} d^d\chi\, (-2\nu)\phi_+\delta\phi_-
+ \int_{A_-} d^d\chi\, (+2\nu)\phi_-\delta\phi_+ \,.
\ee
The source in $A_+$ is $J_+ = \phi_-\vert_{A_+}$, while in $A_-$ the source is $J_-=-\phi_+\vert_{A_-}$.
As before, $\phi_+$ and $\phi_-$ are determined everywhere determined by these sources:
\begin{align}
\phi_+(\chi\in A_+) &= 
\int_{A_+}d^d\chi'\, [K^+]_{\Delta_+}(\chi;\chi') J_+(\chi')
-\int_{A_-}d^d\chi'\, [K^-]_{\Delta_+}(\chi;\chi') J_-(\chi')
\,,
\end{align}
and similarly for $\phi_-$.
Here we see $a_{J_-}=-1$ appearing again in the second term.

Let us use this to find the two-point function $\cG_{++}(\chi,\chi')$ for $\chi,\chi'\in A_+$.
Assume we only have a source in $A_+$, so that $J_-=0$.
The expression for the variation of the on-shell action tells us that now
\be
\delta (-S_\text{os}) = \int_{A_+} d^d\chi\, (-2\nu)\phi_+ \delta J_+ 
\,,
\ee
giving
\be
\corr{\varphi_+(\chi)}_{J_+} = \frac{\delta (-S_\text{os})}{\delta J_+(\chi)} = 2\nu\,\phi_+(\chi) 
\,,
\ee
and hence
\be
\cG_{++}(\chi,\chi') 
= \corr{\varphi(\chi)\varphi(\chi')}
= 2\nu\frac{\delta\phi_+(\chi')}{\delta J_+(\chi)}
= 2\nu \, [K^+]_{\Delta_+}(\chi;\chi') \,.
\ee
Replication of this procedure yields the three independent two-point functions:
for $\chi_\pm,\chi_\pm'\in A_\pm$,
\begin{subequations}
\label{eq:correlators}
\begin{align}
\cG_{++}(\chi_+,\chi'_+) &= +2\nu [K^+]_{\Delta_+}(\chi_+;\chi_+') \label{eq:G++}
\\
\cG_{--}(\chi_-,\chi_-') &= -2\nu [K^-]_{\Delta_-}(\chi_-;\chi_-') \\
\cG_{+-}(\chi_+,\chi_-') &
= -2\nu [K^-]_{\Delta_+}(\chi_+;\chi_-')
= +2\nu [K^+]_{\Delta_-}(\chi_-';\chi_+)
\label{eq:G+-}
\,. 
\end{align}
\end{subequations}
Note that \req{eq:correlators} is invariant under ($+\leftrightarrow -$, $\nu\to-\nu$, $\chi\leftrightarrow \chi'$), as it should be.

\subsection{Interface fusion and other generalizations}
\label{sec:defect fusion}
As we have emphasized, the mixed boundary value problem approach 
can deal with more general geometries than the Janus approach.
Let us take a moment to touch on a geometry relevant to a topic of 
particular interest for the theory of conformal interfaces: \emph{interface fusion}.

The methods discussed above can be used to understand the fusion 
properties of two double trace interfaces with the opposite 
orientation.
As a simple example, consider the case of two concentric spherical 
interfaces with opposite orientations, corresponding to $\cft_-$ on
region $A_-$, which is interrupted by an annular region $A_+ = \{ x\,|\,R_1<|x|<R_2\}$ of $\cft_+$.
This configuration preserves $\SO(d)$ symmetry.

The Green's function is obtained using the tools outlined in this 
section (a detailed example is worked out in 
appendix~\ref{sec:K as mbvp}): expand the bulk-boundary propagator 
(or the Green's function) using spherical wave solutions of the 
bulk wave equation.
The region where we impose condition~\ref{K condition 2} is 
different from that of appendix~\ref{sec:K as mbvp}, and so the 
ansatz relevant to the spherical interface -- found in 
equation~\req{eq:psi to g} -- must be replaced by an ansatz 
appropriate to the new $A_-$.
Similarly, the analog of~\req{eq:g integral equation}, required to 
satisfy~\ref{K condition 3}, will now give a more complicated 
integral equation that must be solved to obtain $K$.

\medskip
Carrying out this procedure explicitly is complicated, and we leave it for future work.
Configurations with even smaller symmetry groups can in principle 
be considered, but the difficulty of solving the mixed boundary 
value problem increases quickly as the degree of symmetry is 
reduced.

\section{Green's function from harmonic methods}
\label{sec:integral transform}
Let us now turn to the explicit computation of the interface propagators in the case of a spherical interface.
In this section we take the boundary to be spherical, and $A_+$ to be a hemisphere.
The computation is simplest in Janus coordinates on $H^{d+1}$~\cite{Bak:2003jk}, which make the $\SO(d,1)$ symmetry of the defect manifest.
We will mostly use the coordinates
\begin{align}
ds_{H^{d+1}} 
    &= \frac{dz^2}{4z^2(1-z)^2} + \frac{ds^2_{H^d}}{4z(1-z)}
    \qquad
    z\in(0,1)
\,,
\label{eq:z metric}
\end{align}
and reserve $x,x',\ldots$ to refer to points on the $H^d$ slice.
(For a summary of the relationship of these to other useful 
coordinate systems on $H^{d+1}$, see appendix~\ref{sec:coordinate systems}.)
In these coordinates, the boundary is split into two components: $A_+$,
which lies at $z\to 1$, and $A_-$, at $z\to 0$.
The interface lies at the boundary of $H^d$, with the limit taken along 
any surface of constant $z$.

To solve (A,B) we begin by decomposing $G$ with respect to eigenfunctions
of the Laplacian on $H^d$.
We choose a basis $\Psi_s(x)$ for the eigenfunctions,
\be
    -\nabla^2_{H^d}\Psi_s(x) = \lambda_s\Psi_s(x)
    \,,
\ee
indexed by some parameters $s$.
The index set is equipped with a measure $d\mu(s)$, with respect to which $\Psi_s(x)$ satisfies the normalization conditions:
\begin{align}
\int d\mu(s)\,\Psi_s(x)\overline{\Psi_s(x')} &= \delta(x,x') 
    \label{eq:delta resolution}\\
\int d^dx\sqrt{g_{H^d}} \Psi_s(x)\overline{\Psi_{s'}(x)} &= \delta(s,s') \,,
\end{align}
with $\delta(s,s')$ the normalized delta function satisfying $\int d\mu(s)\,\delta(s,s')f(s) = f(s')$.

One explicit basis and its measure are given in detail in appendix~\ref{sec:spectral decomposition}.
This basis picks a point $p$ in $H^d$ and decomposes in spherical waves centered around this point.
In this case, $s=(\sigma,\ell)$ where $\ell$ indexes the spherical harmonics on $S^{d-1}$, and $\sigma=\sigma_s\ge 0$ is defined by
\be
    \lambda_s = \left(\frac{d-1}{2}\right)^2 + \sigma_s^2
    \,.
\ee

Since all the functions we use in this paper involve symmetric 
functions $F(x,x')$ of two variables on $H^d$, it is also useful to have a basis for these functions that are Laplacian eigenfunctions. 
As discussed in detail in appendix~\ref{sec:spectral decomposition}, this is straightforward in the spherical basis:
\be
J_\sigma(x,x') = \sum_\ell \Psi_{\sigma,\ell}(x) \overline{\Psi_{\sigma,\ell}(x')}
\ee
is just such an eigenfunction.
It depends only on the $\SO(d,1)$-invariant cross-ratio $\xi$, 
which in Poincar\'e patch coordinates~\req{eq:Hd pp} on $H^d$ is $\frac{(x-x')^2}{4yy'}$.
It further satisfies the useful identity
\be
\int_0^\infty d\sigma\, J_\sigma(x,x') = \delta(x,x') \,.
\ee

A basis for the functions on $H^d$ in hand, our first task is to find the general solution to the wave equation on $H^{d+1}$ adapted to the Janus decomposition.

\subsection{\texorpdfstring%
    {Wave equation on $H^{d+1}$}
    {Wave equation on hyperbolic space}
}
We use the metric \req{eq:z metric}. 
Performing separation of variables with respect to the Janus slicing, we
look for solutions to the wave equation
\be
(-\nabla_{H^{d+1}}^2 + m^2)\phi = 0
\ee
of the form $\phi(z,x) = \Phi(z)\Psi_{s}(x)$.
This gives
\be
    \left\{
        -[z(1-z)]^{d/2+1}\frac{d}{dz}\frac{4}{[z(1-z)]^{d/2-1}}\frac{d}{dz}
        + 4z(1-z)\lambda_s + m^2
    \right\}
    \Phi(z) = 0 \,.
\label{eq:z differential equation}
\ee
The space of solutions is two-dimensional, but in what follows we will be interested in four different solutions:
\begin{align}
\Phi_L^{\pm}(\sigma|z) &= 
[4z(1-z)]^{\Delta_\pm/2}
\hg{\frac{1}{2}\pm\nu+i\sigma}{\frac{1}{2}\pm\nu-i\sigma}{1\pm\nu}{z} 
\label{eq:phiL} \\
\Phi_R^{\pm}(\sigma|z) &= 
[4z(1-z)]^{\Delta_\pm/2}
\hg{\frac{1}{2}\pm\nu+i\sigma}{\frac{1}{2}\pm\nu-i\sigma}{1\pm\nu}{1-z} \,.
\label{eq:phiR}
\end{align}
$\Phi^\pm_{L,R}(\sigma|z)$ have the property that as we approach the left boundary
($z\to 0$), 
\be
\Phi_L^{\pm}(\sigma|z) \sim z^{\Delta_\pm/2} 
\qquad\mathrm{as}\qquad z\to 0 \,,
\ee
while as we approach the right boundary ($z\to 1$),
\be
\Phi_R^\pm(\sigma|z) \sim (1-z)^{\Delta_\pm/2} 
\qquad\mathrm{as}\qquad z\to 1 \,.
\ee
Therefore, $\Phi_L^\pm$ and $\Phi_R^\pm$ give bases with definite 
asymptotics $z^{\Delta_\pm/2}$ on left- and right-hand boundaries, respectively.

Having identified a basis of solutions, we can decompose any solution to the wave equation in the form
\be
f(z,x) = 
\sum_{a=\pm}\int d\mu(s)\, 
g^a_{L,R}(s) \,
\Phi_{L,R}^a(\sigma_s|z) \,
\Psi(s|x) \,.
\ee
We are free to choose as we like whether to expand in terms of 
$\Phi_L^\pm$ or $\Phi_R^\pm$.
Note that when it will cause no confusion, we will frequently 
abbreviate $\Phi^+_L(\sigma|z)$ by $\Phi^+_L(z)$, and so forth.

\subsubsection*{Connection coefficients}
In what follows, we will need the linear transformation between the bases $\Phi_L^\pm$ and $\Phi_R^\pm$.
This is given by Kummer's connection formulae~\req{eq:hg z->1-z}:
\begin{align}
&&
\Phi_{L}^a &= \sum_{b=\pm} A^{ab}\Phi_R^b &
\Phi_{R}^a &= \sum_{b=\pm} A^{ab}\Phi_L^b 
&&
\label{eq:connection}
\end{align}
with 
\begin{align}
A^{\pm\pm} &= \mp\frac{\cosh(\pi\sigma)}{\sin(\pi\nu)} &
A^{\pm\mp} &= 2^{\pm 2\nu}
\frac{\Gamma(1\pm\nu)\Gamma(\pm\nu)}
	{\Gamma(\frac{1}{2}\pm\nu+i\sigma)\Gamma(\frac{1}{2}\pm\nu-i\sigma)} \,.
\label{eq:connection coefficients}
\end{align}
Note that the connection coefficients are symmetric under the exchange of $L\leftrightarrow R$.
Applying the change of basis twice implies the consistency relation
\begin{align}
A^{\pm\pm}A^{\pm\pm} + A^{\pm\mp}A^{\mp\pm} &= 1 \\
A^{\pm\pm}A^{\pm\mp} + A^{\pm\mp}A^{\mp\mp} &= 0 \,.
\label{eq:connection conistency}
\end{align}

\subsection{Green's function}
We are now in a position to decompose the Green's function with respect to the functions $\Psi_s$ and $\Phi_{R,L}^\pm$.
Actually, there are four linearly independent Green's functions $G^{ab}$ with $a,b=\pm$:
\be
G^{ab}(X,X') \sim 
\left\{
\begin{array}{ll}
g_1(x,X')z^{\Delta_a/2} + O(z^{\Delta_a/2+1}) \qquad &z\to 0 \\
g_2(x,X')(1-z)^{\Delta_b/2} + O([1-z]^{\Delta_b/2+1})\qquad & z\to 1
\end{array}
\right\}\,.
\ee
Thus the standard Green's function $G^{++}$ has $\Delta_+$ asymptotics 
on both boundary components, while that with $\Delta_-$ asymptotics on the left boundary and $\Delta_+$ asymptotics on the right boundary is $G^{-+}$.

Any Green's function satisfies the condition
\be
(-\nabla_{H^{d+1}}^2 + m^2)G(X,X') = \delta(X,X') \,,
\label{gf def}
\ee
where $\delta(X,X')$ is the covariant delta function on $H^{d+1}$.
We begin with an ansatz for $G^{ab}$ in terms of eigenfunctions on $H^d$,
\be
G^{ab}(X,X') = \int d\mu(s)\,
\Psi_s(x)\overline{\Psi_s(x')}
\left\{\begin{array}{ll}
\displaystyle
A_{s}(z')\,\Phi_L^a(\sigma_s | z) \quad z<z' \\
\displaystyle
B_{s}(z')\,\Phi_R^b(\sigma_s | z) \quad z>z' 
\end{array}\right\} 
\,.
\ee
Applying $(-\nabla^2_{H^{d+1}}+m^2)$ and using the resolution 
\req{eq:delta resolution} of the delta function on $H^d$, 
\req{gf def} becomes the condition
\be
A(z)\p_z\Phi^a_L(z) - 
B(z)\p_z\Phi^b_R(z)
= 2^{d-1}[z(1-z)]^{\frac{d}{2}-1} 
\ee
which is solved by 
\be
A(z) = 
2^{d-1}[z(1-z)]^{\frac{d}{2}-1} 
\frac{\Phi^a_R(z)}{\cW[\Phi_R^b,\Phi^a_L](z)} \,,
\qquad
B(z) = 
2^{d-1}[z(1-z)]^{\frac{d}{2}-1} 
\frac{\Phi^b_L(z)}{\cW[\Phi_R^b,\Phi^a_L](z)} \,,
\ee
with $\cW[f,g](z)=f\p_zg-g\p_zf$ the Wronskian.

Define $W^{ab}_{MN} = \cW[\Phi^a_M,\Phi^b_N]$, and
$w_{MN}^{ab}$ by $W_{MN}^{ab}=w_{MN}^{ab}W_{RR}^{+-}$.
The Wronskians are found from 
\be
W_{RR}^{+-} = 2^{d}\nu[z(1-z)]^{\frac{d}{2}-1}
\ee
together with the values for the connection coefficients
\begin{align}
w_{LR}^{+-} &= A^{++} &
w_{LR}^{-+} &= -A^{--} &
w_{LL}^{+-} &= -1 \\
w_{LR}^{--} &= A^{-+} &
w_{LR}^{++} &= -A^{+-} 
\,,
\end{align}
all others being determined by $w_{NM}^{ba}=-w_{MN}^{ab}$.
This gives the final form for the Green's function:
\begin{align}
G^{ab}(X,X') 
&= \frac{1}{2\nu}
\int d\mu(s)\,
    \cA^{ab}_{\sigma_s}
    \Psi_s(x) \overline{\Psi_s(x')}
    \left\{\begin{array}{ll}
        \displaystyle \Phi_L^a(\sigma_s ; z\phantom{'}) \Phi_R^b(\sigma_s ; z') \quad& z<z' \\
        \displaystyle \Phi_L^a(\sigma_s ; z') \Phi_R^b(\sigma_s ; z\phantom{'}) & z>z' 
    \end{array}\right\} 
    \\
&= \frac{1}{2\nu}
\int d\sigma\, \cA^{ab}_\sigma J_\sigma(X,X') 
    \left\{\begin{array}{ll}
        \displaystyle \Phi_L^a(\sigma ; z\phantom{'}) \Phi_R^b(\sigma ; z') \quad& z<z' \\
        \displaystyle \Phi_L^a(\sigma ; z') \Phi_R^b(\sigma ; z\phantom{'}) & z>z' \end{array}\right\} \,.
    \label{gf}
\end{align}
with $\cA^{ab}_\sigma = \frac{1}{w^{ba}_{RL}(\sigma)}$ explicitly given by
\begin{align}
    \cA_\sigma^{++} &= A^{+-} &
    \cA_\sigma^{+-} &= A^{--} \\
    \cA_\sigma^{--} &= -A^{-+} &
    \cA_\sigma^{-+} &= -A^{++}
\,.
\end{align}

\subsection{Green's function and the bulk-boundary propagator}
\label{sec:K}
Consider now the bulk-boundary propagator, which is obtained from the Green's function as follows: if $\rho$ is a defining function on $H^{d+1}$, then
\be
K(\rho,x;x') = -\frac{1}{2\Delta-d}\lim_{\rho'\to 0}\frac{1}{\rho'^{\Delta}}G(\rho,x;\rho',x') \,,
\ee
where $\Delta$ is the scaling dimension of the operator living at the boundary point $x'$.
In the coordinate system~\req{eq:z metric} (and in the conformal frame such that the boundary metric is $H^d$), the defining function is $\rho=\sqrt{4z(1-z)}$.

Alternatively, the bulk-boundary propagator can be characterized by \ref{K condition 1}-\ref{K condition 3}.
Denote by $K^{ab}_M(X;x')$ ($M=L,R$) the bulk-boundary propagator for the
$(ab)$ interface, with insertion at the point $x'$ on boundary $M$. 
In the notation of section~\ref{sec:dti} this means, for example, that $K^- = K^{-+}_L$ and $K^+ = K^{-+}_R$. 
We give expressions for $K_L^{ab}$; the generalization to $K_R^{ab}$ is obvious.
In the Janus conformal frame the conditions become
\begin{enumerate}
\item $(-\nabla^2_{\hyp^{d+1}}+m^2)K^{ab}_L(X,x')=0$.
\item The coefficient of $z^{\Delta_{-a}/2}$ near the left-hand boundary $z\to 0$ is the covariant delta function $\delta(x,x')$.
\item The coefficient of $(1-z)^{\Delta_{-b}/2}$ near the right-hand boundary $z\to 1$ vanishes.
\end{enumerate}
The first and third properties imply that $K$ can be expanded in the form
\be
K(X;x') = \int d\mu(s) \,
    \kappa_s(x')
    \Psi_s(x)
    \Phi_{R}^{b}(\sigma_s | z)  
\,.
\ee
To impose the second property, we use the connection relations
$\Phi_R^b = A^{b+}\Phi_L^+ + A^{b-}\Phi_L^-$, together with the 
fact that $\Phi_L^{\pm}=\#\,z^{\Delta_\pm/2}+O(z^{\Delta_\pm/2+1})$ as
$z\to 0$.
If we are to get the covariant delta function, the coefficient of 
this term must give the resolution of the delta function \req{eq:delta resolution}, implying
\be
\kappa_s(x') = \frac{1}{A^{b,-a}}\overline{\Psi_s(x')} \,.
\ee
Hence,
\be
    K^{ab}_L(X;x') 
    = \int d\mu(s) \frac{1}{A^{b,-a}} \Psi_s(x) \overline{\Psi_s(x')}
    \Phi^b_R(\sigma_s | z) \,.
\label{eq:Kab-integral}
\ee

A simple example is given by the standard bulk-boundary propagator with insertion on the left boundary, 
$K_{\Delta_+}=K^{++}_L$.
In the spherical basis, $\int d\mu(s)=\int d\sigma\sum_\ell$. 
Carrying out the sum over $\ell$ gives
\begin{multline}
K_{\Delta_+}(z,x;x') = 
[4z(1-z)]^{\Delta_+/2} \times \\
\int_0^\infty d\sigma\,
\frac{\nu}{4^\nu}\left|
	\frac{\Gamma(\half+\nu+i\sigma)}{\Gamma(1+\nu)}
\right|^2
J_\sigma(x,x')
\,
\hg{\frac{1}{2}+\nu+i\sigma}{\frac{1}{2}+\nu-i\sigma}{1+\nu}{1-z} \,.
\end{multline}
This integral can be evaluated straightforwardly by expanding in a power series in $(1-z)$ and using the integral identity%
\footnote{This is derived by applying the Olevskii transform to $(1+x)^{-a-b}$.}
\be
\frac{1}{\pi}\int_0^\infty ds \left|
	\frac{\Gamma(a+is)\Gamma(b+is)\Gamma(c+is)}{\Gamma(2is)}
\right|^2 \hg{a+is}{a-is}{a+c}{{-x}}
= \frac{\Gamma(a+b)\Gamma(a+c)\Gamma(b+c)}{(1+x)^{a+b}}
\,.
\label{eq:olevskii-gamma}
\ee
Summing the series in $(1-z)$, we find
\be
K_{\Delta_+}(z,x;x') = \frac{\Gamma(\Delta_+)}{\pi^{d/2}\Gamma(\nu)}
\left(\frac{2(z+\xi^2)}{\sqrt{z(1-z)}}\right)^{-\Delta_+}
= \frac{\Gamma(\Delta_+)}{\pi^{d/2}\Gamma(\nu)}(4\Xi^2)^{-\Delta_+}
\label{eq:K++}
\ee
with 
\be
\Xi^2 = \lim_{z'\to 0}\sqrt{4z'(1-z')} \chi_{d+1}^2(X,X')
= \frac{\xi(x,x')+z}{\sqrt{4z(1-z)} }
\,.
\ee
Equation~\req{eq:K++} is related, as it should be, by a Weyl 
transformation to the usual Poincar\'e patch expression.
This can be seen by noting that $\Xi^2$ is a conformal covariant 
factor associated to our choice of defining functional, $\sqrt{4z(1-z)}$. 
The corresponding object for the Poincar\'e patch is
\be
\Xi_\text{p.p.}^2(u,\vec x;\vec x') 
= \lim_{u'\to 0}u' \frac{(\vec x-\vec x')^2 + (u-u')^2}{4uu'}
= \frac{1}{4}\frac{(\vec x-\vec x')^2 + u^2}{u} \,.
\ee
Replacing $\Xi$ by $\Xi_\text{p.p.}$ in~\req{eq:K++} gives the 
standard bulk-boundary propagator.

\subsubsection*{Interface bulk-boundary propagator \texorpdfstring{$K^{+-}_L$}{}}
The bulk-boundary propagator for a non-trivial defect is found in the same way.
Equation~\req{eq:Kab-integral} now takes the form
\be
K_L^{+-}(z,x;x') = \frac{\sin\pi\nu}{\pi}\int_0^\infty d\sigma 
\left|\Gamma(\tfrac{1}{2}+i\sigma)\right|^2 J_\sigma(\chi_d^2) \Phi_R^-(\sigma \,|\, z) \,.
\ee
Using Euler's transformation we can write
\be
\Phi_R^-(\sigma\,|\,z)=z^{\Delta_+/2}[4(1-z)]^{\Delta_-/2}
\hg{\tfrac{1}{2}+i\sigma}{\tfrac{1}{2}-i\sigma}{1-\nu}{1-z} \,.
\ee
Power expanding in $(1-z)$, the integral can be carried out using~\req{eq:olevskii-gamma}, and summing gives 
\be
K_L^{+-}(z,x;x')=\frac{\sin\pi\nu}{\pi}\frac{\Gamma(d/2)}{\pi^{d/2}}
\frac{[4z(1-z)]^{\Delta_+/2}}{4^{\Delta_+}(1-z)^\nu}(1+\xi)^{-d/2}
\hg{d/2}{1}{1-\nu}{\frac{1-z}{1+\xi}}
\,.
\ee
Using Euler's transformation gives the form
\be
K_L^{+-}(z,x;x')=\frac{\sin\pi\nu}{\pi}\frac{\Gamma(d/2)}{\pi^{d/2}}
(4\Xi^2)^{-\Delta_+}
\left(\frac{\xi+z}{1-z}\right)^{\nu}
\hg{d/2}{-\nu}{1-\nu}{-\frac{1-z}{\xi+z}} 
\ee
which can also be nicely represented as
\be
K_L^{+-} = K^{++}_L \times \frac{\sin\pi\nu}{\pi}
\frac{\Gamma(\nu)\Gamma(d/2)}{\Gamma(\Delta_+)}
\left(\frac{\xi+z}{1-z}\right)^{\nu} 
\hg{d/2}{-\nu}{1-\nu}{-\frac{1-z}{\xi+z}}
\,.
\label{eq:K+-}
\ee

\subsubsection*{Other bulk-boundary propagators}
All other propagators can be obtained from these two using the relations
\be
K_L^{a,b}(z,x;x') = K_R^{b,a}(1-z,x;x')
\qquad
K_M^{a,b}(z,x;x') = K_M^{-a,-b}\bigr\vert_{\nu\mapsto-\nu} 
\,.
\ee

\subsection{Interface operator spectrum}
\label{sec:boundary operators}
One of the key features to understand in any interface CFT is the spectrum of operators living on the defect.
Fortunately, from the holographic point of view there is a simple and elegant way to identify the interface operators~\cite{Aharony:2003qf}.
Say we have a single scalar field $\phi$ which couples only to the background geometry.
If the background corresponds to a conformal interface, $\SO(d,1)$ invariance implies the linearized equation of motion can be written in the form
\be
(-\nabla_{H^d}^2+\cD)\phi(z,x)
\ee
where $x$ is the coordinate on $H^d$, and $\cD$ is a differential operator built using only the transverse coordinate $z$.
If we expand $\phi$ in eigenmodes of the operator $\cD$,
\be
\phi(z,x) = \sum_a \phi_a(x) \psi_a(z)
\qquad
\cD\psi_a(z) = m_a^2\psi_a(z)
\,,
\ee
then $\phi_a(x)$ satisfies the standard scalar field equation on $H^d$ with mass $m_a^2$.
Each $\phi_a$ is now the bulk dual to a defect operator of dimension
\be
\Delta_a = \frac{d-1}{2}+\nu_a \,,
\qquad
\nu_a = \sqrt{\frac{(d-1)^2}{4}+m_a^2} \,.
\ee

For double trace interfaces the analysis is particularly simple, as the equation of motion is simply the standard bulk equation of 
motion in Janus coordinates.
The relevant eigenmodes can be found by making the substitution 
$i\sigma\to\nu_a$ in~\req{eq:phiL} and~\req{eq:phiR}, giving us two
convenient bases for the solution space,
\begin{align}
\psi_{a,L}^{\pm}(z) &= 
[4z(1-z)]^{\Delta_\pm/2}
\hg{\frac{1}{2}\pm\nu+\nu_a}{\frac{1}{2}\pm\nu-\nu_a}{1\pm\nu}{z} 
\label{eq:psiL} \\
\psi_{a,R}^{\pm}(z) &= 
[4z(1-z)]^{\Delta_\pm/2}
\hg{\frac{1}{2}\pm\nu+\nu_a}{\frac{1}{2}\pm\nu-\nu_a}{1\pm\nu}{1-z} 
\,.
\label{eq:shiR}
\end{align}
Our problem now is to identify the allowed values of $\nu_a$.
Let us say that the left boundary has $\Delta_-$ asymptotics, and the right, $\Delta_+$.
An allowed eigenmode must satisfy these same asymptotics, which is 
only possible if $\psi_{a,R}^+$ is proportional to $\psi_{a,L}^-$.
Using the connection coefficients \req{eq:connection coefficients} 
(once again replacing $i\sigma\to\nu_a$), we find that this is true when $\cos(\pi\nu_a)=0$.
Throwing out redundant choices, the allowed values of $\nu_a$ are $\nu_a=\frac{1}{2} + a$ (with $a = 0,1,2,\ldots$), yielding the interface operator spectrum:
\be
\Delta_a = \frac{d}{2} + a \,,
\qquad
a = 0,1,\ldots
\ee

Of course, above we only considered those operators descending from the bulk field $\phi$.
However, at $O(1)$ in the $1/N$ expansion this is the only bulk field modified by the defect. 
Boundary primaries built from other fields simply have dimensions 
of the form $\Delta+n$, with $\Delta$ the dimension of a CFT bulk operator $\cO$;
these operators are merely descendants $\p_y^n\cO$, where $y$ is the coordinate transverse to the interface.
Only at $O(1/N)$ does a generic primary $\cO$ develop singularities as it is brought to the defect, giving rise to a shift in the 
conformal dimension of the corresponding boundary operator.
Of course, there are also the multi-trace operators, whose 
dimensions in the large $N$ limit are simply the sum of the 
dimensions of their component operators.

Finally, note that in the above we have chosen the standard quantization for all operators. 
However, there is one operator which lies in the unitarity window: the operator $\cO_0$ dual to $\phi_0$, which has dimension $\frac{d}{2}$.
The corresponding double trace operator has dimension $d$, matching that of the interface displacement operator \cite{Billo:2016cpy}, 
which can be used to generate deformations in the interface shape. 
This strongly suggests that this double trace operator should be identified with the displacement operator.
Since $\cO_0$ is the leading boundary operator in the expansion of the bulk operator $\varphi$, this is consistent with the 
CFT expectation that the displacement operator takes the form $\#\varphi^2+\cdots$.

\section{Correlation functions}
\label{sec:correlation functions}
With the interface bulk-boundary propagator in hand, we turn now to the computation of CFT observables.
This section will deal with the two-point functions.
Recall that the bulk field $\phi$ is dual to a boundary operator $\varphi_+$ of dimension $\Delta_+$ in $A_+$, and to an operator $\varphi_-$ of dimension $\Delta_-$ in $A_-$.
There are therefore three different correlation functions that we can compute: 
\be
\cG_{ab}(x,x') = \corr{\varphi_a(x)\varphi_b(x')}
\qquad
a,b=\pm, \; x\in A_a,\; x'\in A_b \,.
\ee
We begin in section~\ref{sec:2ptfcn} by deriving explicit 
expressions for these two-point functions from the results of 
sections~\ref{sec:corr} and \ref{sec:K}.  
Section~\ref{sec:conformal blocks} uses the conformal block 
expansion of the two-point function to give an alternate derivation of the spectrum of interface primaries at large $N$.

\subsection{Evaluation of the two-point functions}\label{sec:2ptfcn}
Section~\ref{sec:corr} showed how to extract two-point functions from the bulk-boundary propagators.
This can be done using the closed form expressions of 
section~\ref{sec:integral transform}, and we do so for $\cG_{++}$ and $\cG_{--}$ in section~\ref{sec:ppoverlap}.
It is, however, also instructive to work with the representation 
obtained from solving the dual integral equation, as this approach is more general. 
To illustrate this procedure, we therefore derive $\cG_{-+}$ in 
section~\ref{sec:pmoverlap} using the integral representation of appendix~\ref{sec:K as mbvp}.

\subsubsection{%
\texorpdfstring
    {$\corr{\varphi_+(x)\varphi_+(x')}$ and $\corr{\varphi_-(x)\varphi_-(x')}$}
    {++ and -- correlators}
}\label{sec:ppoverlap}
To evaluate the two-point function $\cG_{++}$ for operator insertions in the $A_+$ region, recall that in the standard holographic normalization, $\cG_{++}=2\nu [K_+]_{\Delta_+}$.
The bulk-boundary propagator in Janus frame was given in equation~\req{eq:K+-}.

We make our computation in Poincar\'e patch coordinates on the $H^d$ slices, $ds_{H^d}^2=\frac{d\vec x^2+dy^2}{y^2}$, corresponding to a planar interface.
We wish to compute the correlation function in a flat conformal frame, which requires including the additional Weyl factor $(yy')^{-\Delta_+}$.
Combining this factor with equation~\req{eq:G++} gives the correlator
\begin{align}
\cG_{++}(x,x') &= \frac{2\nu}{(yy')^{\Delta_+}}\lim_{z\to 0}[4z(1-z)]^{-\Delta_+/2}K^{+-}(z,x;x') \\
&= 
\frac{2\nu}{(4yy')^{\Delta_+}}
\frac{\sin\pi\nu}{\pi}\frac{\Gamma(d/2)}{\pi^{d/2}}
\xi^{-d/2}\hg{d/2}{-\nu}{1-\nu}{-\frac{1}{\xi}} \\
&=
\frac{c(\Delta_+)}{|x-x'|^{2\Delta_+}}\left[1 + \frac{\sin\pi\nu}{\pi}\frac{\Gamma(d/2)\Gamma(\nu+1)}{\Gamma(\Delta_++1)}\xi^{\Delta_+}\hg{d/2}{\Delta_+}{\Delta_++1}{-\xi}
\right] 
\end{align}
where $c(\Delta_+)=\frac{2\nu\Gamma(\Delta_+)}{\pi^{d/2}\Gamma(\nu)}$ is the standard holographic normalization factor for scalar correlators.
For the planar interface, the conformal cross ratio takes the form $\xi=\frac{(x-x')^2}{4yy'}$.

When comparing with CFT we will use the canonically normalized correlation function
\be
G_{++}(x,x') 
= \frac{1}{|x-x'|^{2\Delta_+}}\left[
1 + 
B\,\xi^{\Delta_+}
\hg{\Delta_+}{d/2}{\Delta_++1}{-\xi}
\right]
\label{GreensFunctionOneSide}
\ee
with
\be
B=\frac{\Gamma(d/2)\Gamma(\nu+1)}%
{\Gamma(\Delta_++1)}\frac{\sin\pi\nu}{\pi}\,.
\label{eq:B-coefficient}
\ee
The $\varphi_-\varphi_-$ correlator is obtained from this correlation function by combining the reflection $y\mapsto -y$ together with the replacement $\nu\mapsto -\nu$.

\subsubsection{%
\texorpdfstring%
    {$\corr{\varphi_-(\chi)\varphi_+(\chi')}$}
    {+- correlator}
\label{sec:pmoverlap}
}
We evaluate this propagator using the results of appendix~\ref{sec:K as mbvp}, which are derived in Poincar\'e patch coordinates $(u,\chi)$ on $H^{d+1}$. 
The boundary points $\chi$ can be expressed in spherical coordinates with radial coordinate $r$; the interface is located on the sphere $r=R$, and $A_+$ is in the interior region.
The evaluation of this two-point function can be reduced by $\SO(d,1)$ transformation to the case where $r'=0$. 
Equation \req{eq:G+-} tells us we should compute $[K_+]_{\Delta_-}$.
Due to equation~\req{g-function}, as $r'\to 0$ the only harmonic that contributes is $\ell=0$.
We will therefore evaluate the $\ell=0$ contribution for $r'>0$, and then send $r'\to 0$.
(We must perform the process this way: it involves a distributional integral for which the limit does not commute with the integral.)

Set $\ell=0$ and take $r>R$.
We take $Y_0=1$, in which case $c_0 = (\mathrm{vol}\,S^{d-1})^{-1}$.
Inserting \req{eq:g solution} into \req{eq:Kdel-} and using \req{eq:W-S minus} gives
\be
[K_{+,\ell=0}]_{\Delta_-} = c_0\frac{\sin\pi\nu}{\pi}\frac{1}{r^{d-2}}\int_0^R ds \frac{1}{(r^2-s^2)^{1-\nu}}\frac{d}{ds}\left[\frac{\theta(s-r')}{(s^2-r'^2)^\nu}\right]
\,.
\ee
Once we integrate by parts, we can take the limit $r'\to 0$ to obtain
\be
[K_{+,\ell=0}]_{\Delta_-} = 2 c_0\frac{\sin\pi\nu}{\pi}\frac{1}{r^d}\left(\frac{r^2}{R^2}-1\right)^\nu
\,,
\ee
and using equation \req{eq:G+-} gives us the correlator itself,
\be
\cG_{-+}(\chi,0)
=
\corr{\varphi_-(\chi)\varphi_+(0)}
=
\frac{c_{+-}}{r^{\Delta_-}R^{2\nu}}\left(1-\frac{R^2}{r^2}\right)^\nu 
\qquad
c_{+-} = 2\nu \frac{\sin\pi\nu}{\pi}\frac{\Gamma(d/2)}{\pi^{d/2}}
\,.
\label{eq:phi+- non-cov}
\ee

Now, $\SO(d,1)$ invariance imples that the two-point function at general $\chi'$ can be written in the form
\footnote{For a planar defect, the prefactor takes the more familiar form $(-2y)^{-\Delta_-}(2y')^{-\Delta_+}$.}
\be
\cG_{-+}(\chi,\chi')
=
\left(\frac{R}{r^2-R^2}\right)^{\Delta_-} \left(\frac{R}{R^2-r'^2}\right)^{\Delta_+}
f(\xi) 
\label{eq:phi+- cov}
\ee
with the conformal cross ratio for a spherical defect given by
\be
\xi(\chi,\chi') = \frac{R^2(\chi-\chi')^2}{(R^2-r^2)(R^2-r'^2)} \,.
\label{eq:xi}
\ee
At $\chi'=0$, $\xi=\frac{r^2}{R^2-r^2}$, so $\frac{r^2}{R^2}=\frac{\xi}{1+\xi}$.
We can find $[K_+]_{\Delta_-}$ at general values of $\chi'$ simply by making
this replacement in the above expression.
(Note that when $r'<R<r$, $\xi<-1$.)
Setting $r'=0$ and equating \req{eq:phi+- cov} and \req{eq:phi+- non-cov} gives 
\be
f(\xi) = c_{+-}(-\xi)^{-d/2}
\,.
\ee
The correlator thus becomes
\be
\corr{\varphi_-(x)\varphi_+(x')}
=
c_{+-}
\left(\frac{R}{r^2-R^2}\right)^{\Delta_-} \left(\frac{R}{R^2-r'^2}\right)^{\Delta_+}
(-\xi)^{d/2}
=
\frac{c_{+-}}{|x-x'|^d}\left(\frac{r^2-R^2}{R^2-r'^2}\right)^{\nu}
\,.
\ee
If we perform a conformal transformation to planar interface 
coordinates $x=(\vec x,y)$ such that $\Delta_+$ is the region given by $y>0$, the correlator takes the form
\be\label{pmoverlap-planar}
\corr{\varphi_-(x)\varphi_+(x')}
=
\frac{c_{+-}}{|x-x'|^d}\left(\frac{-y}{y'}\right)^{\nu}
=
\frac{c_{+-}}{(-2y)^{\Delta_-}(2y')^{\Delta_+}}(-\xi)^{-d/2} \,,
\ee
where now $\xi = \frac{(x-x')^2}{4yy'}$.
For some purposes it is useful to work with the folded picture correlator $\hat \cG_{-+}$.
With $\hat x=(\vec x,-y)=(\vec x,\hat y)$, and $\hat\xi=\frac{(\hat x-x')^2}{4\hat yy'}=-1-\xi$, this is defined by 
\be
\hat\cG_{-+}(\hat x,x')
=
\corr{\varphi_-(\hat x)\varphi_+(x')}_\text{folded}
=
\corr{\varphi_-(x)\varphi_+(x')}
=
\frac{c_{+-}}{(2 \hat y)^{\Delta_-}(2y')^{\Delta_+}}(1+\hat\xi)^{-d/2}
\,.
\label{eq:folded G+-}
\ee
Finally, for comparison with CFT it is useful to give the canonically normalized folded correlator
\be\label{PMoverlapCanonicalNorm}
\hat G^\mathrm{norm.}_{-+}(\hat x,x')
=
\sqrt{\frac{\sin\pi\nu}{\pi\nu}}\frac{\Gamma(d/2)}{\sqrt{\Gamma(\tfrac{d}{2}+\nu)\Gamma(\tfrac{d}{2}-\nu)}}
\frac{(1+\hat\xi)^{-d/2}}{(2\hat y)^{\Delta_-}(2y')^{\Delta_+}}
\,.
\ee

\subsection{Fusion channels and defect spectrum\label{sec:conformal blocks}}
Bulk correlation functions in CFT are well known to be completely determined by the structure coefficients in the theory $C^p_{qr}$. 
If $\varphi_p$ denote the quasi-primary operators of the theory, 
\be
\varphi_p(x) \varphi_{p'}(x') = \sum_q C^q_{pp'} C[x-x';\p_{x'}] \varphi_q(x')
\ee
holds as an operator equation, where $C[x-x',\p_{x'}]$ are operators depending only on conformal dimension.
Inserting this expansion into correlation functions reduces their 
computation to a knowledge of $C^q_{pp'}$, which are model-dependent, and conformal blocks, which are universal.
The requirement of \emph{crossing symmetry} --- that the answer be 
independent of the order in which OPEs are taken --- puts powerful 
constraints on the spectrum and couplings of a CFT, and underlies 
the recent success of the numerical conformal bootstrap methods initiated in~\cite{Rattazzi:2008pe}. 

Using the folding trick, any interface can be thought of as a boundary of the product CFT.
In the presence of a planar boundary any primary $\varphi_p$ has the boundary OPE
\be
\varphi_p(x) = \sum_a B^a_p D[y;\p_{\vec x}] \psi_a(\vec x)
\ee
where $\psi_a$ runs over the $\SO(d,1)$ quasi-primaries living on the boundary, and $D$ is a function depending only on the dimension $\Delta_a$.
Here we have decomposed $x=(\vec x,y)$, with $y$ the distance to the boundary.
In the presence of an interface, this expansion can be used to evaluate any bulk object in terms of interface correlators.
In particular, interface two-point functions can be decomposed in 
terms of boundary conformal blocks, which were first derived in~\cite{McAvity:1995zd}. 
The requirement that this process yields the same result as the bulk OPE imposes constraints on the CFT and its boundary.

In the presence of an interface, non-trivial constraints arise 
already at the level of two-point functions, and so the structure 
implied by the bulk and boundary OPEs should be realized in the two-point functions $\cG_{ab}$. 
Since at leading order in the $1/N$ expansion double trace 
interfaces do not see coupling to any other fields, the conformal 
block structure at this order should only involve operators 
realized holographically in terms of the field $\phi$ itself.
We will show in this section that the operator dimensions predicted by the conformal block decomposition of the two-point functions 
match those derived in section~\ref{sec:boundary operators}, and so indeed satisfy this condition.
Furthermore, we use our results to derive relations between OPE 
coefficients, which we will compare in specific cases to known CFT results in section~\ref{sec:cft}.

In what follows we work with the canonically normalized correlation functiosn $G_{ab}$.

\subsubsection{Bulk fusion channel}
We begin with the bulk fusion channel, derived from the OPE as $\xi\to 0$. 
The correlator of two scalar bulk operators $O$ and $O'$ has the bulk conformal block decomposition~\cite{McAvity:1995zd}
\be
\corr{O(x)O'(x')}_\cD
= 
\frac{1}{(2y)^{\Delta}(2y')^{\Delta'}}\xi^{-(\Delta+\Delta')/2}\sum_{q} C_{OO'}^q B_q^\mathrm{id} 
\cF(\Delta_q,\Delta-\Delta'|\xi) 
\ee
where $q$ runs over bulk quasi-primaries, and the bulk channel conformal block is
\be
\cF(\Delta,\delta\,\vert\,\xi) 
= \xi^{\Delta/2}
\hg{\tfrac{1}{2}(\Delta+\delta)}{\tfrac{1}{2}(\Delta-\delta)}{\Delta+1-\tfrac{d}{2}}{-\xi}
\,.
\ee
When the argument $\delta=0$ we simply omit it.
In the case of 2d CFT this is the expression for the \emph{global} conformal block; these are the only blocks that will be visible in our decomposition even in 2d CFT, since Virasoro blocks degenerate to global conformal blocks at large central charge.

\medskip\noindent
$G_{++}$:
The bulk fusion channel is obtained from an inspection of \req{GreensFunctionOneSide}.
The first term corresponds to the identity block, while the leading behavior of the second term corresponds to an operator of dimension $2\Delta_+$.
A closed form for the conformal block decomposition of the second term follows from the formulae of appendix~\ref{sec:hg sum}, 
\be
x^{\Delta_+} \hg{d/2}{\Delta_+}{\Delta_++1}{-x}
= 
\sum_{n=0}^\infty 
\frac{(\nu)_n(\nu+1)_n(\Delta_+)_n}
{n!(\Delta_++1)_n(\Delta_++\nu+n)_n}
\cF(2\Delta_++2n\,\vert\, x)
\,.
\ee
Therefore the $\varphi_+\varphi_+$ OPE contains a quasiprimary $\cO_n$ with non-vanishing one point function for every dimension $\Delta_n=2\Delta_++2n$ $(n=0,1,\ldots)$.
This result has a straightforward interpretation: the only 
operators contributing to the exchange channel at this level are double trace operators built from the descendants of $\varphi_+$.
Such an interpretation is consistent with the fact that the interface is built from only one bulk field $\Phi$.
We can be much more precise: at leading order in the $1/N$ expansion, the OPE coefficients satisfy the relation
\be
C^n_{\varphi_+\varphi_+}B_n^\text{id} = 
\frac{\sin\pi\nu}{\pi}\frac{\Gamma(d/2)\Gamma(1+\nu+n)}{\Gamma(\Delta_++1+n)}\frac{(\nu)_n(\Delta_+)_n}{(\Delta_++\nu+n)_n}
\,.
\ee
The same analysis applies to $G_{--}$ under $\nu\to-\nu$.

\bigskip\noindent
$G_{-+}$: To apply the BCFT formulae we work with a planar interface in the folded picture on the upper half plane.
Write 
\be
\hat G_{-+}(x,x')
= c'
\frac{\hat\xi^{-d/2}}{(2y)^{\Delta_-}(2y')^{\Delta_+}}\times 
\hat\xi^{d/2}(1+\hat\xi)^{-d/2}
\ee
with $c'=\sqrt{\frac{\sin\pi\nu}{\pi\nu}}\frac{\Gamma(d/2)}{\sqrt{\Gamma(\Delta_+)\Gamma(\Delta_-)}}$, 
so that
\be
c'\hat\xi^{d/2}(1+\hat\xi)^{-d/2} = \sum_q C^q_{\varphi_-\varphi_+}B_q^\text{id} \cF(\Delta_q,2\nu\,\vert\,\hat\xi) \,.
\ee
Applying~\req{eq:hg sum} with $a=\frac{d}{2}$, $b=c$, $\alpha=\Delta_+$, $\beta=\delta_-$, $\gamma=\tfrac{d}{2}+1$ gives the decomposition 
\be
\hat\xi^{d/2}(1+\hat\xi)^{-d/2}=\sum_{n=0}^\infty \frac{(\Delta_+)_n(\Delta_-)_n}{n!\,(\tfrac{d}{2}+n)_n} {}_3F_2\biggl( \!\begin{array}{c} \tfrac{d}{2}, \tfrac{d}{2}+n, -n \\ \Delta_+, \Delta_- \end{array} \Big|\, 1 \biggr)
\cF(d+2n,2\nu\,|\,\hat\xi) \,.
\ee
This implies that there is a contribution from fusion channels containing operators $O_n$ of dimension $\Delta_n=d+2n$.

These have a quite transparent interpretation in terms of the $\varphi_-\varphi_+$ OPE: since $\varphi_-$ and $\varphi_+$ live in different sectors of the product CFT their OPE is non-singular, and
clearly closes in terms of the double trace operators built from descendants of $\varphi_-$ and $\varphi_+$.
In particular, we can read off the coefficient product
\be
C^{O_n}_{\varphi_-\varphi_+}B_{O_n}^\text{id}
=
c'
 \,.
\ee
Obviously, the operator $O_0$ can simply be chosen as the 
normal-ordered coincidence limit $O_0=(\varphi_+\varphi_-) - \text{(divergence)}$.
In this normalization, 
\be
B^\text{id}_{O_0}=c'
\,.
\ee

\subsubsection{Boundary fusion channel}
The bulk-boundary OPE
\be
\cO(x) = \sum_a B^a_O D[y,\p_{\vec x}] \psi_a(\vec x) 
\ee
allows bulk operators to be expanded in terms of boundary primary 
operators $\psi_a$ and their descendants, which we take to be orthogonal
\be
\corr{\psi_a(\vec x)\psi_b(\vec x')} = \frac{\cN_a \delta_{ab}}{|x-x'|^{\Delta_a}}
\,.
\ee
Inserting this OPE into a two-point function, one can derive the representation~\cite{McAvity:1995zd}
\be
\corr{O(x)O'(x')}=\frac{1}{(2y)^{\Delta}(2y')^{\Delta'}}
\sum_{a}\cN_a B_O^a B_{O'}^a 
\cF_\p(\Delta_a \,\vert\, \xi)
\,,
\ee
where the boundary channel conformal block $\cF_\p$ is given by
\be
\cF_\p(\Delta\,\vert\,\xi) 
= \xi^{-\Delta}\,
\hg{\Delta}{\Delta-\tfrac{d}{2}+1}{2\Delta-d+2}{-\frac{1}{\xi}} \,.
\label{eq:bdy decomp}
\ee

\bigskip\noindent
$G_{++}$:
Using the hypergeometric indentity \req{eq:hg z->1/z}, we can write
\begin{align}
G_{++}(x,x') 
&= 
\frac{1}{(4yy')^{\Delta_+}}\frac{\sin\pi\nu}{\pi}\frac{\Gamma(d/2)\Gamma(\nu)}{\Gamma(\Delta_+)}\xi^{-d/2}\,
\hg{d/2}{-\nu}{1-\nu}{-\frac{1}{\xi}}
\,,
\end{align}
which is in the appropriate form to apply~\req{eq:bdy decomp}.
The decomposition follows from the results of 
appendix~\req{sec:hg sum} and takes the form
\be
G(x,x')=
\frac{1}{(4yy')^{\Delta_+}}\frac{\sin\pi\nu}{\pi}\frac{\Gamma(d/2)\Gamma(\nu)}{\Gamma(\Delta_+)}
\sum_{k=0}^\infty
\frac{k!(d/2)_k(1+\nu)_k}{(2k)!(1-\nu)_k}\cF_\p(\tfrac{d}{2}+k\,\vert\,\xi)
\ee
so that we have a contribution from a pair of boundary operators of dimension $\frac{d}{2}+k$ for each $k=0,1,\ldots$.
This is the same as the boundary operator spectrum found in section~\ref{sec:boundary operators}.
Note that as we approach the boundary, the dominant contribution comes from a boundary operator $\psi$ of dimension $\frac{d}{2}$,
\be
\varphi_+(x) \sim \frac{1}{(2y)^{\nu}} \psi_0(\vec x) + \cdots
\ee
where
\be
\corr{\psi_0(\vec x)\psi_0(\vec x')}_\cD = \frac{\sin\pi\nu}{\pi\nu}
\frac{\Gamma(d/2)\Gamma(\nu)}{\Gamma(\Delta_+)}
\frac{1}{|\vec x-\vec x'|^{d}} \,.
\ee
As discussed in section~\ref{sec:boundary operators}, it is very natural to guess that $\psi\psi$ fuses into the displacement operator, 
\be
D(x)\sim {:}\psi\psi{:}(x)
\,,
\ee
which has dimension $d$.
In particular, we expect that the displacement operator two-point 
function is determined at leading order by the $\phi\phi\phi\phi$ four-point function.

\bigskip\noindent
$G_{-+}$:
Write the folded picture correlator
\be
\hat G_{-+}(x,x') = 
\frac{c'}{(2y)^{\Delta_-}(2y')^{\Delta_+}}
\frac{1}{\hat\xi^{d/2}}(1+\hat\xi^{-1})^{-d/2}
= 
\frac{1}{(2y)^{\Delta_-}(2y')^{\Delta_+}}
\sum_a
B_{\varphi_-}^a B_{\varphi_+}^a \cN_a
\cF_\p(\Delta_a\,|\,\hat\xi)
\,.
\ee
Setting $a=\tfrac{d}{2}$, $b=c$ in~\req{eq:hg sum} we have
\be
(1+\hat\xi)^{-d/2}=\sum_{k=0}^\infty \frac{(d/2)_k(1)_k}{(2k)!}\hat\xi^{-k}\hg{1+k}{-k}{1}{1}\hg{\tfrac{d}{2}+k}{1+k}{2+2k}{-\hat\xi^{-1}}
\,.
\ee
The first hypergeometric function can be evaluated by replacing the ``$c$'' parameter 1 by $1+\epsilon$, using Gauss' summation formula, and taking the limit $\epsilon\to 0$, giving $(-)^k$.
We therefore obtain
\be
(1+\hat\xi)^{-d/2}=\sum_{k=0}^\infty \frac{k!\,(d/2)_k}{(2k)!}(-)^k \cF_\p(\Delta_k\,|\,\hat\xi) 
\qquad
\Delta_k = \tfrac{d}{2}+k
\ee
matching the spectrum derived in~\ref{sec:boundary operators}.
The fusion coefficients satisfy 
\be
B_{\varphi_+}^k B_{\varphi_-}^k \cN_k
= 
c' (-)^k \frac{k!\,(d/2)_k}{(2k)!} \,.
\ee
%

\section{Interface partition function%
\label{sec:boundary entropy}}
We now turn to the computation of the simplest quantum effect of double
trace interfaces: the leading contribution to the sphere free 
energy due to a double trace interface on the equator, at large $N$.
In the specific case $d=2$, this quantitiy coincides with the boundary entropy, or $g$ factor~\cite{Affleck:1991tk}, of 2d CFT.
The defect free energy is the leading non-extensive contribution to the 
thermal free energy in the expansion in $\beta/L$, where $\beta^{-1}$ is the 
temperature and $L$ is the length of a very long semi-infinite cylinder.
Thus, for example in 2d BCFT one can write
\be
\log Z = \frac{c}{12}\frac{L}{\beta} + \log g + O(\beta/L) \,.
\ee

Computing the overall one-loop correction to the free energy requires both UV and IR regulators.
The defect contribution to the free energy, however, can be 
expressed as the difference of two free energies defined using the same UV regulator, which is a UV finite quantity. 
Our construction is as follows.
Take the bulk theory to be $\cft_+\otimes \cft_-$.
Into this theory we can introduce the double trace interface joining $\cft_+$ on the left to $\cft_-$ on the right, and \emph{vice versa}.
Consider the difference $\Delta F$ of the free energy of this theory with the defect, $F_{+\otimes -}^\cD$, and without the defect, $F_{+\otimes -}$.
The bulk contribution to the free energy cancels between these two terms, and so we have
\be
F_\mathrm{defect} = \frac{1}{2}\Delta_F = \frac{1}{2}(F_{+-} + F_{-+} - F_{++} - F_{--}) \,.
\ee
Here $F_{ab}$ denotes the free energy of a theory with a single copy of $CFT_a$ on the left and $CFT_b$ on the right.
The $g$ factor is given by
\be
g^2 = \frac{Z_{CFT_+\otimes CFT_-}^\text{defect}}{Z_{CFT_+\otimes CFT_-}^\text{vacuum}}
= \left(\frac{[\det \scD]^{++}[\det\scD]^{--}}{[\det \scD]^{+-}[\det\scD]^{-+}}\right)^{-1/2}
\ee
with $[\det\scD]^{ab}$ the functional determinant of $\scD=(-\Box+m^2)$ with $(a,b)$ boundary conditions.
Using
\be
\frac{d}{dm^2}[\tr\log\scD]^{ab} = \int d^{d+1}\!X\sqrt{g_{H^{d+1}}}\, G^{ab}(X,X)
\ee
and $\nu^2=\frac{d^2}{4}+m^2$ we find
\be
\frac{d}{d\nu}\log g^2 = -\nu \int d^{d+1}\!X\sqrt{g_{H^{d+1}}}\, \scH(X)
\label{eq:dg/dnu}
\ee
where 
\be
\scH(X) = \lim_{X'\to X}\Bigl(
    G^{+-}(X,X') + G^{-+}(X,X') - G^{++}(X,X') - G^{--}(X,X')
\Bigr)
\,.
\ee
Since when $\nu=0$ the defect is trivial (and hence $g=1$), the value of $g$ is given by the integral $\log g^2=\int_0^\nu d\nu' \frac{d}{d\nu '}\log g^2$.

\subsection{Regulator}
Equation~\req{eq:dg/dnu} is infrared divergent and must be regularized by cutting off the bulk integral.
Expressing the metric in the form
\be
ds^2_{H^{d+1}} = d\rho^2 + \sinh^2\!\rho\,(d\theta^2 + d\Omega_{d-1}^2) \,,
\ee
we choose the cutoff surface defined by $\rho=\rho_*$, which corresponds to computing the CFT partititon function on the sphere.
To compute the one-loop contribution of the interface, we need to express the cutoff surface in Janus coordinates~\cite{Bak:2016rpn}.
For our purposes the coordinate system
\be
ds^2_{H^{d+1}} = \frac{d\tau^2}{4\tau^2(1-\tau)}+\frac{1}{\tau}ds_{H^d}^2
\,, \qquad
\tau = 4z(1-z)
\ee
is useful; 
note however that the function $\tau(z)$ is 2-to-1 and symmetric about $z=1/2$.
Writing the metric on $H^d$ in the form
\be
ds^2_{H^d} = \frac{dw^2}{w(1+w)} + 4w(1+w)d\Omega_{d-1}^2 \,,
\ee
the Poincar\'e ball coordinates and Janus coordinates are related by
\be
\cosh\rho = \sqrt{\frac{1+2w}{\tau}}
\quad\qquad
\tanh\rho\,\sin\theta = \sqrt{\frac{2w}{1+2w}}
\,.
\ee
The intersection of the cutoff surface with a leaf of given $\tau$ is therefore defined by the relation $w=w_*(\tau)$, where
\be
\tau^{1/2} = \epsilon(1+2w_*) \,,
\qquad
\epsilon = \frac{1}{\cosh\rho_*} \,.
\ee
Note that $w_*\ge 0$, which means that the minimum value of $\tau$ is given by
\be
\tau\ge\tau_* = \epsilon^2 \,.
\ee

\subsection{Sphere free energy and the \texorpdfstring{$g$-factor}{g-factor}}\label{sec:gfactor}
To proceed, we use equation~\req{gf} to write $\scH$ in the form
\begin{multline}
\scH(X) = \frac{1}{2\nu}\int d\sigma \, 
\cN_\sigma\biggl(
\frac{1}{A^{--}}\Phi_L^+(z)\Phi_R^-(z)
- \frac{1}{A^{++}}\Phi_L^-(z)\Phi_R^+(z) \\
- \frac{1}{A^{--}}\Phi_L^+(z)\Phi_R^+(z)
+ \frac{1}{A^{-+}}\Phi_L^-(z)\Phi_R^-(z)
\biggr) 
\end{multline}
where
\be
\cN_\sigma = |\Psi_{\sigma,0}(0)|^2
\,.
\ee
Kummer's formulae~\req{eq:connection} allow us to write this as
\be
\scH(X) = \frac{1}{2}\int_0^\infty d\sigma\,
\cN_\sigma \sum_{s=\pm}
c_s \bigl[\Phi_L^s(z)\bigr]^2 \,,
\qquad\;\;
c_\pm = \pm\frac{\sin\pi\nu}{\cosh\pi\sigma}\frac{1}{4^{\pm\nu}}\left|\frac{\Gamma(\tfrac{1}{2}\pm\nu+i\sigma)}{\Gamma(1\pm\nu)}\right|^2
\,.
\ee

The trace now takes the form
\be
\int d^{d+1}X\sqrt{g_{H^{d+1}}} \scH(X)
= \frac{1}{2}
\int_0^\infty d\sigma\,\cN_\sigma \sum_{s=\pm} c_s \times
\int d^{d+1}X \sqrt{g_{H^{d+1}}} (\Phi_L^s)^2 \,.
\ee
To evaluate the inner integral, note that the integral over $H^d$ 
in Janus coordinates simply gives the regulated volume $(\vol H^d)_*$.
For $z<\tfrac{1}{2}$, a quadratic transformation of ${}_2F_1$ allows us to express $\Phi_L^\pm$ in the form
\be
\Phi_L^\pm(z) = \tau^{\Delta_+/2}\hg{a_\pm}{b_\pm}{c_\pm}{\tau}\,,
\qquad\text{with}\qquad
a_\pm = (b_\pm)^* = \frac{1}{2}(\tfrac{1}{2}\pm\nu+i\sigma) \,,
\quad
c_\pm = 1\pm\nu \,.
\ee
Since the integral is symmetric under $z\mapsto 1-z$, in the above integral we may make the replacement
\be
\int d^{d+1}X \sqrt{g_{H^{d+1}}} (\Phi_L^s)^2
\rightarrow
(\vol H^d)_*\,
2\int_{\tau_*}^1 \frac{d\tau}{2\tau^{d/2+1}\sqrt{1-\tau}}
\tau^{\Delta_s}\left[ \hg{a_s}{b_s}{c_s}{\tau} \right]^2 \,;
\ee
the factor of 2 is required since $\tau$ only covers half the geometry. 
Using the identities
\be
\left[ \hg{a}{b}{a+b+\tfrac{1}{2}}{\tau} \right]^2
= 
\htt{2a}{2b}{a+b}{a+b+\tfrac{1}{2}}{2a+2b}{\tau}
\ee
and 
\begin{multline}
\int_{\epsilon^2}^1 d\tau\,
\tau^{s\nu-1}(1-\tau)^{1/2}
\htt{2a_s}{2b_s}{\tfrac{1}{2}+s\nu}{1+s\nu}{1+2s\nu}{\tau} \\
= \frac{\Gamma(\tfrac{1}{2})\Gamma(s\nu)}{\Gamma(\tfrac{1}{2}+s\nu)}
\htt{2a_s}{2b_s}{s\nu}{1+s\nu}{1+2s\nu}{1}
- \frac{\epsilon^{2s\nu}}{s\nu} + O(\epsilon^{1+s\nu})
\end{multline}
together with the doubling formula for $\Gamma$, this becomes
\be
(\vol H^d)_* \left[
4^{s\nu}
\frac{\Gamma(s\nu)\Gamma(1+s\nu)}{\Gamma(1+2s\nu)}
\htt{2a_s}{2b_s}{s\nu}{1+s\nu}{1+2s\nu}{1}
- \frac{\tau_*^{s\nu}}{s\nu} + \cdots
\right]
\ee
with $\cdots$ vanishing as $\epsilon$ (and thus $\tau_*$) approaches~0.
We obtain
\begin{multline}
\int d^{d+1}\sqrt{g_{H^{d+1}}} \scH(X) = 
(\vol H^d)_*
\sum_s 
\int_0^\infty\!d\sigma\, 
2^{2s\nu-1}\cN_\sigma
c_s
\times \\
\left[
    \frac{\Gamma(s\nu)\Gamma(1+s\nu)}{\Gamma(1+2s\nu)}
    \htt{\half+s\nu+i\sigma}{\half+s\nu-i\sigma}{s\nu}{1+s\nu}{1+2s\nu}{1}
    - \frac{\epsilon^{2s\nu}}{s\nu} + \cdots
\right]
\,.
\end{multline}

This expression has two sources of IR divergence.
The first is from the volume $(\vol H^d)_*$, while the second is due to the term proportional to $\epsilon^{-\nu}$. 
$(\vol H^d)_*$ has an expansion (for $d\not\in 2\N$) in powers $\epsilon_*^{1-d+2m}$, $m=0, 1, 2, \ldots$.
Provided $d$ is not an odd integer, the divergences fall into two 
non-overlapping series, which can presumably be eliminated by counterterms that do not affect the finite part of the trace.
Alternatively, we can define the integral with $s=-$ by analytic continuation to $\nu<0$.
Either way, the $\epsilon^{-2\nu}$ divergence can be dropped, and 
the regularized volume replaced by the standard renormalized hyperbolic volume $(\vol H^d)_\text{ren}$.
We do this from now on.

The $\sigma$ integral now takes the form
\begin{multline}
s\,C_1\frac{\Gamma(s\nu)\Gamma(1+s\nu)}{\Gamma(1+2s\nu)}\frac{1}{\pi}
\int_0^\infty d\sigma
\left|
    \frac{\Gamma(\tfrac{d-1}{2}+i\sigma)\Gamma(\half+i\sigma)^2\Gamma(\half+s\nu+i\sigma)}
    {\Gamma(d/2)\Gamma(1+s\nu)\Gamma(2i\sigma)}
\right|^2 \times \\
\htt{\half+s\nu+i\sigma}{\half+s\nu-i\sigma}{s\nu}{1+s\nu}{1+2s\nu}{1}   
\end{multline}
where $C_1 = \frac{1}{2} \frac{\Gamma(d/2)}{(4\pi)^{d/2}}\frac{\sin\pi\nu}{\pi}$.
We can evaluate the $\sigma$ integral using the results of section~\ref{sec:mellin}.
The renormalized volume integral then takes the form
\be
(\int \scH(X))_\text{ren} = 
C_1 (\vol H^d)_\text{ren}
\sum_{s=\pm} s\;
\frac{\Gamma(\Delta_s)\Gamma(s\nu)\Gamma(1+s\nu)}{\Gamma(\Delta_s+1)\Gamma(1+2s\nu)}
\htt{\Delta_s}{s\nu}{1+s\nu}{\Delta_s+1}{1+2s\nu}{1} \,.
\label{eq:Hren}
\ee

Equation~\req{eq:Hren} can be evaluated using the 3-term ${}_3F_2$ relation~\req{eq:3-term relation} given in the appendix.
Combining this with \req{eq:Hren} and \req{eq:dg/dnu} gives the value 
\be
\frac{d}{d\nu}\log g^2 = 
-\frac{\nu\,\cos\pi\nu}{(4\pi)^{d/2}}
\frac{\Gamma(\Delta_+)\Gamma(\Delta_-)}{\Gamma(\frac{d}{2}+1)} \times
(\vol H^d)_\text{ren} \,.
\ee
Under dimensional regularization the volume of $H^d$ becomes \cite{Diaz:2007an}
\be
(\vol H^d)_\text{ren} = \pi^{\frac{d-1}{2}} \Gamma\big({-\tfrac{d-1}{2}}\big)
\,,
\ee
so that
\be
\frac{d}{d\nu}\log g^2 = - \nu\, \frac{\cos\pi\nu}{\cos\tfrac{\pi d}{2}} \frac{\Gamma(\Delta_+)\Gamma(\Delta_-)}{\Gamma(1+d)} \,.
\ee

It is interesting to compare this to the value of the difference between the renormalized action of $\cft_+$ and $\cft_-$ \cite{Diaz:2007an}:
\be
\frac{d}{d\nu}(S_{\cft_+} - S_{\cft_-}) = \nu\, \frac{\sin\pi\nu}{\sin\tfrac{\pi d}{2}}
\frac{\Gamma(\Delta_+)\Gamma(\Delta_-)}{\Gamma(1+d)} \,,
\ee
from which one can extract the shift in central charge. 
It is amusing to speculate that the similarity of these expression 
may indicate some deeper relation between the change in central 
charge under RG flow, and the $g$ factor for the corresponding RG defect.

Note that our result diverges as $d$ approaches odd integers, 
corresponding to a logarithmic divergence with respect to $\epsilon$.
This reflects the fact that in odd dimensions, the defect free 
energy is associated to a conformal anomaly localized on the 
interface locus~\cite{Herzog:2015ioa,Fursaev:2015wpa,Solodukhin:2015eca}.

\subsubsection*{Explicit values}
As examples, we give explicit expressions in several cases where the $g$ factor has no ambiguities.

\noindent
$\mathbf{d=2:}$
\be
\frac{d}{d\nu}\log g^2 = \frac{\pi}{2} \nu^2\cot\pi\nu
\label{eq:g-d=2}
\ee

\noindent
$\mathbf{d=4:}$
\be
\frac{d}{d\nu}\log g^2 = -\frac{\pi}{4!} \nu^2(1-\nu)^2\cot\pi\nu
\ee

\noindent
$\mathbf{d=6:}$
\be
\frac{d}{d\nu}\log g^2 = \frac{\pi}{6!} \nu^2(1-\nu)^2(2-\nu)^2\cot\pi\nu
\ee

\section{Comparison to field theory results\label{sec:cft}}

In this section we check our bulk results against computations we 
make directly in the CFT. 
We are interested in particular in the coefficients
appearing in the correlation functions of 
section~\ref{sec:2ptfcn}, and in the $g$ factor of 
section~\ref{sec:gfactor}. 
We will compute two-point functions for small $\nu$ by means of 
conformal perturbation theory in section~\ref{sec:confpert}, and 
show they coincide at large $N$ with the results of section~\ref{sec:2ptfcn}. 
We further calculate the $g$ factor and several overlaps of the 
solvable RG interfaces constructed in $d=2$ coset models by Gaiotto in~\cite{Gaiotto:2012np}.
We will show in section~\ref{sec:d2CFT} that, assuming the higher 
spin/$\cW$-CFT correspondence of~\cite{Gaberdiel:2010pz}, these 
coincide at large $N$ with our bulk results in two dimensions for all values \mbox{$0\leq\nu<1$}.

\subsection{Coefficients from conformal perturbation theory}\label{sec:confpert}
A check of the coefficients appearing in the correlation functions 
of section~\ref{sec:2ptfcn} can be made against conformal perturbation theory. 
A CFT can be perturbed by adding a term 
\begin{equation}
\delta{S}=\kappa\, \epsilon^{d-\Delta_\cO}\int 
d^dx\,
\cO(x) \,+\,S_{\rm c.t.}
\end{equation}
to the Euclidean action, where $\cO$ is an operator of conformal dimension $\Delta_\cO$, 
$\kappa$ is a dimensionless coupling constant, and $\epsilon$
is a (scheme-dependent) length scale which we will take
to be a position space short-distance cutoff. $S_{\rm c.t.}$ is the counterterm action arising during the renormalization procedure.
Correlation functions of (renormalized) local 
operators~${\cal O}_i$ of the perturbed CFT 
can be expressed schematically in terms of the correlation functions of the CFT as
\begin{equation}
\cor{\cO_1(x_1)\cdots \cO_n(x_n)}_{\rm pert}=
\frac{\cor{\cO_1(x_1)\cdots \cO_n(x_n)e^{-\delta S}}}%
{\cor{e^{-\delta S}}}\,.
\end{equation} 
For short flows, the right-hand side can be expanded in powers of the renormalized coupling constants.

We are interested in deforming by an operator of the form $\varphi_-^2$, the normal-ordered product of $\varphi_-$ with 
itself, in the case where $\Delta_{\varphi_-}=\frac{d}{2}-\nu$ with $0\le\nu<1$. 
When $\nu=0$ the interface is trivial, while small values of $\nu$ give rise to short RG flows. 
If the CFT has a weakly curved bulk dual, and if $\varphi_-$ is 
dual to a bulk scalar appearing in the path integral, then in the 
large~$N$ limit $\varphi_-$ is a ``generalized free field'' 
(see~\cite{ElShowk:2011ag} and reference~\cite{Jost:1965} therein).
This means that correlation functions factorize into two-point functions by Wick contraction. 
The conformal dimension of this operator is then given by twice the dimension~$\Delta_-$ of $\varphi_-$, making $\varphi_-^2$ a 
marginally relevant operator for small values of $\nu$. 
In the large~$N$ limit it is also expected that $\varphi_-^2$ is 
the only non-trivial relevant operator in the OPE of $\varphi_-^2$ with itself. 
Denote the coefficient of $\varphi_-^2$ in this OPE by $C$. 
In the OPE (position-space cut-off) scheme, the beta function corresponding to~$\kappa$ of the double trace deformation reads
\begin{equation}
\beta=(d-2\Delta_-)\kappa-\tfrac{1}{2}\sphere{d-1}\,C\,
\kappa^2\,+\,\mathcal{O}(\kappa^3)\,,
\end{equation}
where $\sphere{d-1}=\frac{2\pi^{\frac{d}{2}}}{\Gamma(\frac{d}{2})}$ is the volume of $S^d$.
The value of $\kappa$ at the IR fixed point (where $\beta=0$) is therefore
\begin{equation}\label{IRcoupling}
\kappa=\frac{4\nu}{\sphere{d-1}\,C}\,,
\end{equation}
such that perturbative results in $\kappa$
correspond to perturbative results in $\nu$.

Let us consider a planar interface. Like in 
section~\ref{sec:ppoverlap} we will use the coordinates 
$x=(\vec{x},y)$ but work in the flat conformal frame. Recall that in section~\ref{sec:ppoverlap} we compute the correlation 
function for two scalar insertions $\varphi_-$ at 
points $x$ and $x'$, whose distance from the interface is
denoted $y$ and $y'$.
To first order in $\kappa$, this correlation function is perturbatively given by
\begin{equation}\label{original}
\cor{\varphi_-(x)\varphi_-(x')}_{\rm pert}=
\cor{\varphi_-(x)\varphi_-(x')}
-\kappa\epsilon^{2\Delta_--d}\int_{y''<0} \!\!\! d^dx''
\cor{\varphi^2_-(x'')\varphi_-(x)\varphi_-(x')}\,.
\end{equation}
The integral runs over the half-space $y''<0$,
which does not include the two points $x$ and $x'$.
Conformal invariance allows us to take both $x$ and $x'$ to lie on the positive $y$ axis. 
The correlator inside the integral has the form
\begin{equation}\label{threepointfunction}
\cor{\varphi^2_-(x'')\varphi_-(x)\varphi_-(x')}=
C'|x''-x|^{-2\Delta_-}|x''-x'|^{-2\Delta_-}\,,
\end{equation}
so that the right-hand side of~\eqref{original} is proportional to the integral 
\begin{equation}\label{Id}
I=\int_{y''<0} d^dx''\,|x''-x|^{-d}|x''-x'|^{-d}\,.
\end{equation}
Using spherical coordinates parallel to the interface, 
and $z=-y''$, we have
\begin{align}
I=\int_{0}^\infty dz\int dr\, d\Omega_{d-2}\,r^{d-2}%
(r^2+(z+y)^2)^{-\frac{d}{2}}(r^2+(z+y')^2)^{-\frac{d}{2}}\,.
\end{align}
The angular integral yields the volume $\cA_{d-2}$ of $S^{d-2}$, while the integral over $r$ is of the form
\begin{equation}\label{Id1}
\int_0^\infty dr\, r^{d-2}(r^2+a^2)^{-\frac{d}{2}}
(r^2+b^2)^{-\frac{d}{2}}=\frac{\sphere{d-1}}%
{2\,\sphere{d-2}}\,\frac{(a+b)^{1-d}}{a\,b}\,,
\end{equation}
valid for $a,b > 0$. For the remaining integral over $z$ we use
\begin{equation}\label{Id2}
\int_0^\infty dz\, \frac{2z+y+y'}{(2z+y+y')^d(z+y)(z+y')}%
=\int_0^\infty \frac{dz}{(2z+y+y')^d(z+y)}%
\;+\;\{y\leftrightarrow y'\}\,,
\end{equation}
with 
\begin{equation}\label{Id3}
\int_0^\infty \frac{dz}{(2z+y+y')^d(z+y)} = %
\frac{1}{(2y)^d\,d}\,
\hg{d}{d}{d+1}{-\frac{y'-y}{2y}}\,.
\end{equation}
Using \eqref{Id1}, \eqref{Id2} and \eqref{Id3}, \eqref{Id} is
\begin{align}
I &=
\frac{\sphere{d-1}}{2d}
\left[\frac{1}{(2y)^d}\,\hg{d}{d}{d+1}{-\frac{y'-y}{2y}} 
    + \{y\leftrightarrow y'\}\right] \nonumber\\
&= \frac{1}{d}\frac{\sphere{d-1}}{(4yy')^{d/2}}
\hg{d/2}{d/2}{\tfrac{d}{2}+1}{-\frac{(y'-y)^2}{4yy'}}\,.
\label{Isol}
\end{align}
Combining \eqref{IRcoupling}, \eqref{threepointfunction}, \eqref{Isol}, 
and restoring the $x$ and $x'$ dependence, \eqref{original} becomes
\begin{equation}
\cor{\varphi_-(x)\varphi_-(x')}_{\rm pert}=
|x-x'|^{-2\Delta_-}\left(1\,-\,\frac{\nu}{d}\frac{C'}{C}\frac{4}{(4yy')^\frac{d}{2}}\,
\,
\hg{d/2}{d/2}{\tfrac{d}{2}+1}{-\xi}\right)\,,
\end{equation}
where $\xi=\frac{(x-x')^2}{4yy'}$ is the conformal cross ratio.
To first order in $\nu$, this formula coincides with the one obtained
in section~\ref{sec:ppoverlap}, which was
\begin{equation}
G_{--}=\frac{1}{|x-x'|^{2\Delta_-}}\left[1+B\xi^{\Delta_-} \hg{\Delta_-}{d/2}{\Delta_-+1}{-\xi}\right]\,,\quad
B=-\frac{\Gamma(d/2)\Gamma(1-\nu)}{\Gamma(\Delta_-+1)}\frac{\sin\pi\nu}{\pi}\,,
\end{equation}
provided that
\begin{equation}\label{CCprimeCondition}
C=2C'\,.
\end{equation}
This relation is due to the fact that at leading order, $\varphi_-$ is a generalized free field.
Using Wick contraction it is simple to verify that~\req{CCprimeCondition} is satisfied.
Let us illustrate this in the context of the large-$N$ 
free/Wilson-Fisher interface of the $O(N)$ vector model in $d$ dimensions.%
\footnote{The RG interface between the $O(N)$ free and Wilson-Fisher critical points
for finite $N$ was investigated
in~\cite{Gliozzi:2015qsa}.} 
The theory contains~$N$ scalar fields $\phi_1,\,\ldots,\,\phi_N$. 
The scalar field $\varphi_-$, corresponding up to 
normalization to the operator $\phi_i\phi_i$,
and the double trace operator $\varphi_-^2$, 
corresponding to $(\phi_i\phi_i)^2$, have 
the OPEs
\begin{align}
\varphi_-(x)\varphi_-(0)&=x^{-2\Delta_-}+
\sqrt{\tfrac{2}{N}}x^{-\Delta_-}\varphi_-(0)
+\tfrac{\sqrt{2(N^2-N+3)}}{N}\varphi^2_-(0)+\ldots\,,
\nonumber\\
\varphi^2_-(x)\varphi^2_-(0)&=x^{-4\Delta_-}+
\tfrac{4\sqrt{2N}(N+2)}{N^2-N+3}x^{-3\Delta_-}\varphi_-(0)
+\tfrac{\sqrt{8}(N+8)}{\sqrt{N^2-N+3}}x^{-2\Delta_-}
\varphi^2_-(0)+\ldots\,,
\end{align} 
where ellipses stand for omitted 
irrelevant operators. For $N\rightarrow\infty$,
we obtain
\begin{equation}
C'=\sqrt{2}\,,\qquad C=2\sqrt{2}\,,
\end{equation}
in agreement with~\eqref{CCprimeCondition}.

The two-point function of $\varphi_+$ can be obtained, to first order
in perturbation theory, in a manner analogous to the one just described by perturbing the IR
action with the marginally irrelevant operator 
$\varphi^2_+$. 
If the analogous conditions apply for the OPEs of $\varphi_+$ and $\varphi_+^2$, we indeed obtain the result~\eqref{GreensFunctionOneSide}
(and~\eqref{eq:B-coefficient}).

\bigskip
To compute the perturbative overlap 
across the interface, which we will compare with
section~\ref{sec:pmoverlap}, we start with two
insertions of the operator $\varphi_-$ on the
$y$ axis at positions $y'>0$ and $-y<0$. Let
the perturbation run over the half space $\{ x''\,|\, y''>0$\}, so that
\begin{equation}\label{originalpm}
\cor{\varphi_-(-y)\varphi_-(y')}_{\rm pert}=
\cor{\varphi_-(-y)\varphi_-(y')}
-\kappa\epsilon^{2\Delta_--d}\int_{y''>0} \!\!\! d^dx''
\cor{\varphi^2_-(x'')\varphi_-(-y)\varphi_-(y')}\,.
\end{equation}
This time we need to cut off the integral over $x''$
at radius $\epsilon$ away from $y'$. In order to compute
this, let us split the integral into two parts:
one where the coordinate $y''$ is outside of the slab
$s_\epsilon=(y'-\epsilon,y'+\epsilon)$, and one where it
is inside the slab.

Outside of the slab we have to compute the integral
\begin{align}
I_{out}&=\int_{y''\notin s_\epsilon}\!\!\!\!\! 
d^dx''\cor{\varphi^2_-(x'')\varphi_-(-y)\varphi_-(y')}=\nonumber\\
&=C'\sphere{d-2}\int_{y''\notin s_\epsilon}\!\!\!\! dy''
\int dr\, r^{d-2}((y''+y)^2+r^2)^{-\tfrac{d}{2}}
((y''-y')^2+r^2)^{-\tfrac{d}{2}}\,,
\end{align}
for which we use \eqref{Id1} again to obtain
\begin{equation}
I_{out}=\frac{C'}{2}\,\sphere{d-1}\left[
\int_0^{y'-\epsilon}dy''\,\frac{(y+y')^{1-d}}{(y+y'')(y'-y'')}
+\int_{y'+\epsilon}^\infty dy''\,\frac{(2y''+y-y')^{1-d}}{(y''+y)(y''-y')}
\right]\,.
\end{equation}
The first integral in the square brackets evaluates to
\begin{equation}
\int_0^{y'-\epsilon}dy''\,\frac{(y+y')^{1-d}}{(y+y'')(y'-y'')}=
\frac{1}{(y+y')^d}\left(\log\frac{y'}{y}-\log\frac{\epsilon}{y+y'-\epsilon}\right)\,.
\end{equation}
In the other integral we can split the integrand and shift $y''$, such that
\begin{equation}
\int_{y'+\epsilon}^\infty dy''\,\frac{(2y''+y-y')^{1-d}}{(y''+y)(y''-y')}
=
\int_{\epsilon}^\infty\left( \frac{dz}{(2z+y+y')^dz}+\frac{dz}{(2z+y+y')^d(z+y+y')}\right)\,.
\end{equation}
Using the analogue of~\eqref{Id3} one has
\begin{align}
\int_{\epsilon}^\infty\frac{dz}{(2z+y+y')^dz}&=
\frac{1}{(2\epsilon)^d d}\,
\hg{d}{d}{d+1}{-\frac{y+y'}{2\epsilon}}\nonumber\\
&=
-\tfrac{1}{(y+y')^d}\left(\log\frac{\epsilon}{y+y'}+\log2+\psi(d)+\gamma\right)+\mathcal{O}(\epsilon)\,,
\end{align}
where $\psi$ is the digamma function and $\gamma$ is Euler's constant, together with
\begin{align}
\int_{\epsilon}^\infty\frac{dz}{(2z+y+y')^d(z+y+y')}&=\frac{1}{(2(y+y'))^dd}\,\hg{d}{d}{d+1}{\frac{1}{2}}\nonumber\\
&=\frac{1}{2(y+y')^d}\left(\psi(\tfrac{d+1}{2})-\psi(\tfrac{d}{2})\right)\,,
\end{align}
such that the contribution from outside the slab becomes
\begin{equation}
I_{out}=\frac{C'\sphere{d-1}}{2(y+y')^d}\left(
\log\frac{y'}{y}-2\log\frac{\epsilon}{y+y'}-
\log2+\tfrac{1}{2}\psi(\tfrac{d+1}{2})-
\tfrac{1}{2}\psi(\tfrac{d}{2})-\psi(d)-\gamma
\right)\,.
\end{equation}
Inside the slab we must cut off the integral over the directions
parallel to the interface at an appropriate distance from the $y$
axis, depending on the value of $y''$. Rescaling integration
variables by $y+y'$, we have
\begin{align}
I_\text{in}&=\int_{y''\in s_\epsilon}\!\!\!\!\! 
d^dx''\cor{\varphi^2_-(x'')\varphi_-(-y)\varphi_-(y')}=\nonumber\\
&=\frac{C'\sphere{d-2}}{(y+y')^d}\int_{-\epsilon'}^{\epsilon'}d\eta
\int_{\sqrt{(\epsilon')^2-\eta^2}}^\infty dr\, r^{d-2}
(\eta^2+r^2)^{-\tfrac{d}{2}}
((\eta+1)^2+r^2)^{-\tfrac{d}{2}}\,,
\end{align}
where $\eta$ is the rescaled $y''$, and $\epsilon'$ is
the rescaled cut-off. As $\eta$ is very small,
we can expand the last factor of the integrand.
All odd powers of $\eta$ will drop out in the
integration, 
so that we can write
\begin{equation}
I_\text{in}=\frac{2C'\sphere{d-2}}{(y+y')^d}\int_0^{\epsilon'}d\eta
\int_{\sqrt{(\epsilon')^2-\eta^2}}^\infty\frac{r^{d-2}\,dr}{
(\eta^2+r^2)^{\tfrac{d}{2}}
(1+r^2)^{\tfrac{d}{2}}}\left(1+\mathcal{O}(\eta^2)\right)\,.
\end{equation} 
Changing coordinates to $\tau^2=\eta^2+r^2$ and expanding
the factor $(1+r^2)=(1+\tau^2-\eta^2)$ in $\eta$ again, this
expression can be written as
\begin{equation}
I_\text{in}=\frac{2C'\sphere{d-2}}{(y+y')^d}\int_0^{\epsilon'}d\eta
\int_{\epsilon'}^\infty d\tau\, \tau^{-2}(1-\tfrac{\eta^2}{\tau^2})^{\frac{d-3}{2}}
(\tau^2+1)^{-\frac{d}{2}}\left(1+\mathcal{O}(\eta^2)\right)\,.
\end{equation}
We now employ the binomial series 
\begin{equation}
(1-\tfrac{\eta^2}{\tau^2})^{\frac{d-3}{2}}=\sum_{k=0}^\infty \binom{\frac{d-3}{2}}{k}(-1)^k\frac{\eta^{2k}}{\tau^{2k}}\,,
\end{equation}
which is valid on the domain of integration. 
Note that the $\eta$ integral of the $k^\text{th}$ term of the sum 
yields a suppression by ${\epsilon'}^{\,2k+1}$, while its leading 
contribution to the $\tau$ integral is
\begin{align}
\int_{\epsilon'}^\infty \frac{d\tau}{\tau^{2+2k}(\tau^2+1)^{\frac{d}{2}}}
&=\frac{(\epsilon')^{-(d+2k+1)}}{(d+2k+1)}\,
\hg{d/2}{d/2+k}{\frac{d+3}{2}+k}{-\frac{1}{(\epsilon')^2}}
\nonumber\\
&=\frac{1}{(\epsilon')^{2k+1}}\left(\frac{1}{2k+1}+\mathcal{O}(\epsilon')\right)\,.
\end{align}
We therefore find
\begin{equation}
I_\text{in}=\frac{2C'\sphere{d-2}}{(y+y')^d}\sum_{k=0}^\infty
\binom{\frac{d-3}{2}}{k}\frac{(-1)^k}{(2k+1)^2}\,+\,\mathcal{O}(\epsilon')\,.
\end{equation}
For the sum we have
\begin{equation}
\sum_{k=0}^\infty\binom{\frac{d-3}{2}}{k}\frac{(-1)^k}{(2k+1)^2}=
\frac{\sqrt{\pi}}{4}\,\frac{\Gamma(\frac{d-1}{2})}{\Gamma(\frac{d}{2})}\,%
\big(\gamma+\log4+\psi(\tfrac{d}{2})\big)\,,
\end{equation}
and thus
\begin{equation}
I_\text{in}=\frac{C'\sphere{d-1}}{2(y+y')^d}\left(\gamma+\log4+\psi(\tfrac{d}{2})\right)
\,+\,\mathcal{O}(\epsilon')\,.
\end{equation}
Combining the contributions $I_{out}$ and $I_\text{in}$,
and using the identity
\begin{equation}
\psi(d)-\tfrac{1}{2}\psi(\tfrac{d+1}{2})-\tfrac{1}{2}\psi(\tfrac{d}{2})=\log2
\end{equation}
together with the value~\eqref{IRcoupling} of the coupling constant in the IR, the 
value of the perturbed correlation function~\eqref{originalpm} becomes
\begin{equation}
\cor{\varphi_-(-y)\varphi_-(y')}_{\rm pert}=\frac{1}{(y+y')^d}+
\frac{2C'}{C}\,\frac{\nu}{(y+y')^d}\left(2\log\frac{\epsilon}{(y+y')}%
+\log\frac{y}{y'}\right)\,.
\end{equation}
This expression still contains a divergence in $\epsilon$,
which is eliminated by an appropriate counterterm.
Conformal invariance dictates that to first order in $\nu$, the correlation function must take the form
\begin{equation}\label{pmoverlapgeneralform}
\cor{\varphi_-(-y)\varphi_+(y')}=\frac{f_\nu(\xi)}{y^{\Delta_-}(y')^{\Delta_+}}\,,
\end{equation}
where $\xi=(x-x')^2/(4yy')=(y+y')^2/(4yy')$ is the conformal cross ratio. 
Since the case $\nu=0$ corresponds to the identity interface, the 
function $f_\nu(\xi)$ must satisfy
$f_0(\xi)=\xi^{-\frac{d}{2}}$. Expanding~\eqref{pmoverlapgeneralform}
to first order in $\nu$ and using the dimensions 
$\Delta_\pm=\tfrac{d}{2}\pm\nu$ leads to the condition
\begin{equation}
\frac{2C'}{C}\left(2\log\frac{\epsilon}{y+y'}+\log\frac{y}{y'}\right)+\mathrm{c.t.}
=\frac{(y+y')^d}{(yy')^{\frac{d}{2}}}\,\partial_\nu f_\nu(\xi)\big|_{\nu=0}+\log\frac{y}{y'}\,,
\end{equation}
where ``$\mathrm{c.t.}$'' stands for the counterterm 
contribution. We observe that the condition $C=2C'$ leads to the cancellation of 
the $\log y/y'$ term on both sides. The remaining part
of the left-hand side must then be a function of $\xi$ alone.
In the OPE scheme the counterterm can only depend on the distance
$y+y'$ of the two field insertions, and therefore cannot 
do anything other than precisely eliminate
the logarithmic divergence.
We therefore conclude that
\begin{equation}
\partial_\nu f_\nu(\xi)\big|_{\nu=0}=0\,,
\end{equation}
which makes~\eqref{pmoverlapgeneralform} indeed agree with 
the gravitational result~\eqref{pmoverlap-planar}.

\subsection{Checks from minimal model holography in \texorpdfstring{$d=2$}{d=2}}\label{sec:d2CFT}
In $d=2$, the duality between Vasiliev Higher Spin theory in the 
bulk and Minimal Model CFTs on the boundary belongs to the 
best-understood examples of non-supersymmetric 
holography~\cite{Gaberdiel:2010pz,Gaberdiel:2012uj}. 
The classical bulk contains one massless field of spin $s$ for 
every integer $s\geq 2$, which transform under the higher spin 
algebra~$\hs(\nu)$, depending on (the square of) an \textit{a priori} arbitrary complex number $\nu$.
The theory includes a complex scalar field (with a propagating 
degree of freedom, unlike the topological higher spin fields) 
of mass $m^2=\nu^2-1$, with $-1<\nu<1$.
A single higher spin gravity gives rise to two 
boundary theories: 
under $\nu\mapsto-\nu$ the algebra $\hs(\nu)$ remains unchanged, 
while in the unitary window the scalar field of dimension $\Delta_+$ acquires the alternate quantization~$\Delta_-$. 

The asymptotic quantum symmetry algebra $\cW_\infty(\nu)$ 
associated to $\hs(\nu)$ also arises as the 't Hooft limit of the algebra~$\cW_{N,k}$.
This is the chiral algebra of the CFT $\cM_{N,k}$ based on the coset 
\begin{equation}\label{coset algebra}
\frac{\su(N)_k\otimes \su(N)_1}{\su(N)_{k+1}} \,,
\end{equation}
which has central charge
\begin{equation}
c_{N,k}=(N-1)\frac{k}{N+k}\frac{k+2N+1}{k+N+1}\,.
\end{equation}
The 't Hooft limit takes $N,k\to\infty$ at fixed $\nu=\frac{N}{N+k}$.%
\footnote{For other coset models and their RG flows in the 
't Hooft limit see 
{\it e.g.}~\cite{Ahn:2011pv,Ahn:2012fz}.}
Irreducible representations of the 
coset~\eqref{coset algebra} are labelled by a 
pair~$\Lambda=(\lambda^+,\lambda^-)$ of representation labels
of $\su(N)_k$ and $\su(N)_{k+1}$, respectively.%
\footnote{Here we follow the common convention to suppress an $\su(N)_1$ 
representation label, which is automatically fixed by the choice of
$\lambda^+$ and $\lambda^-$. 
Our conventions for $\su(N)$ can be found in appendix~\ref{SUNconventions}.}
We will only consider charge-conjugate theories with diagonal modular invariant: 
{\it i.e.}, the theory contains only left-right symmetric pairs of 
representations, $\Lambda\otimes\tilde\Lambda$ with $\Lambda\simeq\tilde\Lambda$, and for each such pair in the 
Hilbert space, the charge conjugate pair 
$\overline{\Lambda}\otimes\overline{\tilde\Lambda}$ is present as well.
Despite the left-right symmetry, we write tildes over right-movers for the purpose of clarity.

The large-level limit of such theories is in general a rather 
subtle issue~\cite{Roggenkamp:2003qp,Runkel:2001ng} and leads to 
continuous orbifold 
theories~\cite{Gaberdiel:2011aa,Fredenhagen:2014kia,Gaberdiel:2014cha}.
However, the equivalence with the $\cW_\infty$ algebras in fact holds for 
finite $N$ and $k$ --- and therefore finite $c$ --- since 
an extension of level-rank duality identifies
$\cW_{N,k}\cong \cW_{\infty}(\tfrac{N}{k+N})$ if 
$c=c_{N,k}$~\cite{Altschuler:1990th,Gaberdiel:2012uj}.\\
 
For the unitary theories (where $N$ and $k$ are positive integers) 
there exists a well-known relevant deformation of the 
CFT $\cM_{N,k}$ which has $\cM_{N,k-1}$ as its IR fixed point~\cite{Ahn:1990gn}. 
Gaiotto introduced interfaces corresponding to this RG flow and 
gave a recipe for computing its UV-IR overlaps in~\cite{Gaiotto:2012np}.
The renormalization group flow from $\cM_{N,k}$ to $\cM_{N,k-1}$ 
was proposed in~\cite{Gaberdiel:2010pz} to be the double trace flow from $\cft_-$ to $\cft_+$, and the one-loop computations 
of~\cite{Giombi:2013fka} support this proposal.
Therefore, we expect Gaiotto's interface to be realized holographically as a double trace interface.

Because the bulk scalar field is complex, we must take care to 
include additional factors of 2 when comparing $\log g$ and the overlap coefficients with our bulk computations.

\subsubsection{RG interface construction at finite \texorpdfstring{$N$ and $k$}{N and k}}\label{GaiottoInterfaceConstruction}
Before we come to the results in the 't Hooft limit let us briefly
explain how the interface is 
constructed at
finite (positive integer) $N$ and $k$.
We give the interface
as a boundary condition in the folded theory $\cft_\text{UV}\otimes\overline{\cft}_\text{IR}$.
The chiral algebra of the folded theory is
\begin{equation}\label{doublechiralalg}
\frac{\su(N)_k\otimes \su(N)_1}{\su(N)_{k+1}}\otimes
\frac{\su(N)_{k-1}\otimes \su(N)_{1'}}{\su(N)_{k}}\,,
\end{equation}
where we distinguish the IR copy of the level 1 algebra
from the UV copy by a prime.
The Hilbert spaces of the UV and IR theories decompose into products 
of representations~\mbox{$\Lambda_{i}\otimes \tilde{\Lambda}_{i}$},
where $\Lambda_{i}$ and $\tilde\Lambda_{i}$ ($i=\text{UV, IR}$) are
representations of the left- and right-moving chiral algebra respectively.
The Hilbert space of the folded 
theory will then contain the product of representations
$\Lambda_{UV}\otimes \overline{\tilde{\Lambda}}_{IR}$ for the left-moving 
and $\tilde{\Lambda}_{UV}\otimes\overline{\Lambda}_{IR}$
for the right-moving degrees of freedom.

The boundary condition corresponding to the RG interface 
consists of a projection in the $\su(N)_{k}$ sector, a 
permutation brane in the $\su(N)_1$ sectors, and a Cardy 
state in the sector 
$\su(N)_{k-1}/\su(N)_{k+1}$ 
of~\eqref{doublechiralalg}~\cite{Brunner:2015vva}. 
The projection can be implemented by the topological 
interface~\cite{Gaiotto:2012np}
\begin{equation}\label{defect}
{\cal I}=\sum_{\tilde{\lambda}^+,\lambda^-}\sum_{\lambda}\frac{1}{S^{(k)}_{0\lambda}}\,
{\large \Pi}_{(\lambda,\lambda^-)_{UV}\otimes 
(\tilde{\lambda}^+,\lambda)_{IR}}
\end{equation}
of the product theory. Here $S^{(k)}_{0\lambda}$ is a 
modular $S$ matrix entry of the $\su(N)_k$ WZW model. 
The operators $\Pi$ project onto the subscript representation%
\footnote{Only the left-moving degrees of freedom are indicated here.}
$(\lambda,\lambda^-)_{UV}\otimes 
(\tilde{\lambda}^+,\lambda)_{IR}$, which are the products of UV and IR 
representations sharing a common label $\lambda$ of $\su(N)_k$.
When summed over $\lambda$, these operators implement the 
isomorphism%
\footnote{Notice that two representation labels of 
$\su(N)_1$ are suppressed on both sides in 
equation~\eqref{isomorphism}.}%
~\cite{Crnkovic:1989ug}
\begin{equation}\label{isomorphism}
\{\tilde{\lambda}^+,\lambda^-\}\cong\bigoplus_{\lambda}
(\lambda,\lambda^-)_{UV}\otimes 
(\tilde{\lambda}^+,\lambda)_{IR}\,,
\end{equation}
where the left-hand side denotes a representation of the 
diagonal coset
\begin{equation}
\frac{\su(N)_{k-1}\otimes \su(N)_1\otimes \su(N)_1}{\su(N)_{k+1}}\,.
\end{equation}
In particular, this isomorphism
identifies the $\su(N)_k$
current operators $J^{(k)a}=J^{(k-1)a}+J^{(1')a}$
of the two copies of $\su(N)_k$ in the product theory.

The boundary condition corresponding to the RG interface is 
given by the fusion product of the topological 
interface~${\cal I}$ with the boundary state
\begin{equation}\label{brane}
\braneket{B}=\mathop{{\sum}'}_{\{\tilde{\lambda}^+,\lambda^-\}}
\!\!\!\!
\sqrt{S^{(k-1)}_{0\tilde{\lambda}^+}\bar{S}^{(k+1)}_{0\lambda^-}}
\ishiket{ \{\tilde{\lambda}^+,\lambda^-\} }_{\mathbb{Z}_2}\,.
\end{equation}
The prime in this expression indicates that the sum only runs over 
representations $\{\tilde{\lambda}^+,\lambda^-\}$ where the 
(suppressed) labels 
of the two $\su(N)_1$ parts are identical. The Ishibashi states 
$\ishiket{\{\tilde{\lambda}^+,\lambda^-\}}_{\mathbb{Z}_2}$
are defined such that they implement a permutation (indicated 
by the subscript $\mathbb{Z}_2$) of these $\su(N)_1$ 
parts~\cite{Recknagel:2002qq,Ishikawa:2001zu,Quella:2002ct}.

The prescription for computing the operator overlaps is 
therefore as follows.
Suppose we want to compute the overlap of the UV operator $\Phi^{UV}$ 
and the conjugate of the IR operator $\Phi^{IR}$, which are composed 
of left- and right-moving parts
\begin{equation}
\Phi^{UV}=\phi^{UV}\tilde{\phi}^{UV}\,,\qquad
\Phi^{IR}=\phi^{IR}\tilde{\phi}^{IR}\,.
\end{equation}
The operators
$\phi^{UV}\tilde{\phi}^{IR}$ and 
$\tilde{\phi}^{UV}\phi^{IR}$ then constitute the left- 
and right-moving part of the corresponding operator in 
the doubled theory, respectively. If $\phi^{UV}$ 
is an operator in the representation 
$(\lambda^+,\lambda^-)$, and $\tilde{\phi}^{IR}$ is in 
the representation conjugate to
$(\tilde{\lambda}^+,\tilde{\lambda}^-)$, we write 
$\phi^{UV}\tilde{\phi}^{IR}$ as a state in
the representation $\{\tilde{\lambda}^+,\lambda^-\}$
of the left-hand side of \eqref{isomorphism}.
This image only exists if the representation labels
of $\su(N)_k$ agree, {\it i.e.} if 
$\lambda^+=\tilde{\lambda}^-$.

After the projection we compute the inner 
product of $\phi^{UV}\tilde{\phi}^{IR}$ and the 
$\mathbb{Z}_2$ flipped image of 
$\tilde{\phi}^{UV}\phi^{IR}$, where the latter is obtained by 
exchanging all degrees of freedom of $\su(N)_1$ and 
$\su(N)_{1'}$. This requires that the (suppressed) 
representation labels of $\su(N)_1$ and $\su(N)_{1'}$ 
agree. Finally, the resulting inner product 
must be multiplied with the corresponding 
coefficient of the boundary state, leading to the
formula
\begin{equation}
\cor{\Phi^{UV}\Phi^{IR}|RG}=
\frac{\sqrt{S^{(k-1)}_{0\tilde{\lambda}^+}%
\bar{S}^{(k+1)}_{0\lambda^-}}}{S^{(k)}_{0\lambda}}
\cor{\phi_{(\lambda,\lambda^-)}^{UV}
\tilde{\phi}_{(\tilde{\lambda}^+,\lambda)}^{IR}
\mathbb{Z}_2(\tilde{\phi}_{(\lambda,\lambda^-)}^{UV}
\phi_{(\tilde{\lambda}^+,\lambda)}^{IR})}\,.
\end{equation} 

\subsubsection{The RG interface in the 't Hooft limit}\label{sec:GaiottosRGInterface}

For finite $N$ and $k$, one way to quantify 
the length of an RG flow is to consider the reflectivity
of the RG interface~\cite{Brunner:2015vva}. 
Reflectivity is measured here with respect to specific parts of the
chiral symmetry algebra, and different definitions exist. 
A coefficient which exists for any conformal interface
measures reflection and transmission of energy and 
momentum~\cite{Quella:2006de}. From the matrix
\begin{equation}\label{Rdefinition}
R\,=\,\frac{1}{\cor{0|RG}}
\left(\begin{array}{cc}
\corr{T^{UV}\tilde{T}^{UV}|RG} & 
\Bigl\langle T^{UV}T^{IR}|RG \Bigr\rangle \\
\corr{\tilde{T}^{UV}\tilde{T}^{IR}|RG} & 
\corr{T^{IR}\tilde{T}^{IR}|RG}
\end{array}\right)=:
\left(\begin{array}{cc}
R_{11} & R_{12} \\
R_{21} & R_{22}
\end{array}\right)
\end{equation}
one defines the reflection and transmission 
coefficients
\begin{equation}\label{RTdefinition}
{\cal R}= \mathcal{N}^{-1}(R_{11}+R_{22})\,,\quad
{\cal T}= \mathcal{N}^{-1}(R_{12}+R_{21})\,,
\end{equation}
where 
\begin{equation}
\mathcal{N}=\sum_{i,j}R_{ij}=\frac{c_{N,k}-c_{N,k-1}}{2}\,.
\end{equation}
These coefficients have the property that~${\cal R}+{\cal T}=1$. 
Also, $0\leq {\cal R}\leq 1$ for interfaces between unitary CFTs, 
with ${\cal R}=0$ for topological interfaces and 
${\cal R}=1$ for interfaces which are (totally 
reflective) conformal boundary states.

For our RG interfaces, the matrix $R$ of~\eqref{Rdefinition} is 
rather easy to compute. The (left-moving) energy-momentum tensor 
components of the UV and the IR are given by
\begin{equation}
T^{UV}=T^{(k)}+T^{(1)}-T^{(k+1)}\,,\qquad
T^{IR}=T^{(k-1)}
+T^{(1')}-T^{(k)}\,,
\end{equation}
where 
\begin{equation}
T^{(k)}=\frac{1}{k+N}\sum_{a}:J^{(k)a}J^{(k)a}:
\end{equation}
is the standard Sugawara energy momentum tensor of
the $\su(N)_k$ WZW model.
Following the prescription of 
identifying $J^{(k)a}=J^{(k-1)a}+J^{(1')a}$
and applying the $\mathbb{Z}_2$ transformation
$J^{(1)a}\leftrightarrow J^{(1')a}$
one obtains~\cite{Brunner:2015vva}
\begin{align}\label{CosetR}
R_{11}&\,=\,\frac{N^2-1}{2(k+N)^2}\frac{k+2N+1}{k+N+1}
\,=\,\tfrac{1}{2}\nu^2(1+\nu)+
\mathcal{O}(\tfrac{1}{k},\tfrac{1}{N})\,,
\nonumber\\[6pt]
R_{12}=R_{21}&\,=\,\frac{(N-1)(k-1)(k+2N+1)}%
{2(k+N)^2}\nonumber\\
&\quad\,=\,\tfrac{N}{2}(1-\nu^2)\,
-\,\tfrac{1}{2}(1+\nu^2)\,+\,
\mathcal{O}(\tfrac{1}{k},\tfrac{1}{N})\,,\\[6pt]
R_{22}&\,=\,\frac{N^2-1}{2(k+N)^2}
\frac{k-1}{k+N-1}\,=\,\tfrac{1}{2}\nu^2(1-\nu)
+\mathcal{O}(\tfrac{1}{k},\tfrac{1}{N})\,.\nonumber
\end{align}
We observe that in the 't Hooft limit, the entries $R_{11}$ and $R_{22}$ (related to reflection) remain finite, 
while the off-diagonal entries $R_{12}$ and $R_{21}$ (related to transmission) diverge.
The coefficients ${\cal R}$ and ${\cal T}$, 
however, remain finite, and asymptote to 
${\cal R}=0$ and ${\cal T}=1$,
as for a topological interface.
Notice that in spite of the finite change 
\begin{equation}\label{changeinc}
c_{N,k}-c_{N,k-1}=\frac{2(N-1)N(N+1)}{(N+k-1)(N+k)(N+k+1)}= 
2\nu^3\,+\,\mathcal{O}(k^{-2},N^{-2})
\end{equation}
in central charge, there is in fact no 
contradiction here, 
since the central charges of the UV and the IR theory are both infinite in the 't Hooft limit.
The RG interface in the 't Hooft limit 
is in general not the identity,
as shown by the non-trivial boundary entropy computed in 
section~\ref{sec:gfactor} and confirmed in the next subsection.

One could at this point also compute the overlaps --- 
the matrix $R$ --- for higher spin fields $W_s$ instead of $T$. 
Each higher spin field of the bulk corresponds to a 
descendent of the vacuum representation of the boundary CFT. 
In the coset numerator theory, the state corresponding to the 
field of spin $s$ has the form~\cite{Bouwknegt:1992wg}
\begin{equation}
\ket{W_s}=\sum_{n=0}^sA_n s_{a_1\ldots a_nb_1\ldots b_{s-n}}
J^{(k)a_1}_{-1}\cdots J^{(k)a_n}_{-1}J^{(\ell)b_1}_{-1}\cdots J^{(\ell)b_{s-n}}_{-1}
\ket{0}\,,
\end{equation} 
where $s_{c_1\ldots c_s}$ is proportional to the totally symmetric
invariant tensor of rank $s$ present for $2\leq s\leq N$ in $\su(N)$.
The coefficients $A_n$ are determined by requiring that $\ket{W_s}$
transforms trivially under the denominator 
subalgebra, and by the normalization condition $\langle{W_s}\ket{W_s}=c/s$.
For the example $s=3$ one finds
\begin{align}
&A_0=\eta\,k(k+N)(2k+N) \,,\qquad A_1=-3\eta\,(k+N)(N+1)(2k+N)\,,&\nonumber\\
&A_2=3\eta\,(k+N)(N+1)(N+2)\,,\qquad A_3=-\eta\,(N+1)(N+2)\,,&\\
&\eta=\left(\frac{N}{18(N^2-4)(N+1)^2(N+2)(k+N)^2(2k+N)(k+N+1)^2(2k+3N+2)}\right)^{\frac{1}{2}}\,,&\nonumber
\end{align}
which leads to the overlap matrix
\begin{align}
R_{11}&\,=\,-\frac{(N^2-1)(N+2)(k+2N+1)}{3(k+N)^2(2k+N)(N+k+1)}\,=\, -\frac{\nu^3(\nu+1)}{3(2-\nu)}+\mathcal{O}(\tfrac{1}{k},\tfrac{1}{N})\,,\nonumber\\
R_{12}&\,=\,R_{21}\,=\,\sqrt{\frac{(2k+N-2)(2k+3N+2)}{(2k+N)(2k+3N)}}\frac{(N-1)(k-1)(k+2N+1)}{3(k+N)^2}\\
&\,=\,\tfrac{N}{3}(1-\nu^2)+\frac{3\nu^4-5\nu^2-4}{3(4-\nu^3)}+\mathcal{O}(\tfrac{1}{k},\tfrac{1}{N})\,,\nonumber\\
R_{22}&\,=\,\frac{(N^2-1)(N+2)(k-1)}{3(k+N)^2(2k+3N)(k+N-1)}\,=\,\frac{\nu^3(1-\nu)}{3(2+\nu)}+\mathcal{O}(\tfrac{1}{k},\tfrac{1}{N})\,.\nonumber
\end{align}
The fact that $R_{11}$ is negative for unitary theories is an indication
that the conformal RG interface breaks the higher spin algebra. 
Also, the four entries do not sum up to $(c_{N,k}+c_{N,k-1})/3$, 
that they do not provide a sensible measure of reflection and transmission.

\subsubsection{RG interface boundary entropy}
In the boundary state formalism, the $g$ factor of the RG interface
is the coefficient of the vacuum Ishibashi state in the defect boundary state, 
{\it i.e.},
\begin{equation}\label{gfactor}
g^2=\frac{S_{00}^{(k+1)}S_{00}^{(k-1)}}%
{(S_{00}^{(k)})^2}\,.
\end{equation}
The modular $S$ matrix elements of the right-hand 
side can be found in the standard literature (see 
{\it e.g.}~\cite{DiFrancesco:1997nk}), and are 
reproduced for convenience in appendix~\ref{SUNconventions}. We observe that the $g$~factor can be written as a product
\begin{equation}
g^2={\cal P}_1\,{\cal P}_2
\end{equation}
with
\begin{align}
{\cal P}_1&=\left(\frac{(k+N)^2}{(k+N+1)(k+N-1)}\right)^{\frac{N-1}{2}}\,,\nonumber\\
{\cal P}_2&=\prod_{m=1}^{N-1}\left(\frac{\sin(\frac{\pi m}{k+N+1})\sin(\frac{\pi m}{k+N-1})}{\sin^2(\frac{\pi m}{k+N})}\right)^{N-m}\,.
\end{align}
In the 't Hooft limit, the logarithm of ${\cal P}_1$ only contributes 
at subleading order in $1/N$,
\begin{equation}\label{fulllogP1}
\log{\cal P}_1=-\frac{N-1}{2}\left[\log(1+\tfrac{1}{k+N})+\log(1-\tfrac{1}{k+N})\right]=
\frac{\nu^2}{2N}\,+\,\mathcal{O}(N^{-2})\,.
\end{equation}
In order to compute the logarithm of ${\cal P}_2$, define
\begin{equation}\label{substitutions}
x=\frac{\pi m}{k+N}\,,\qquad \delta x=\frac{\pi}{k+N}\,.
\end{equation}
The following expansion in $\delta x$ holds:
\begin{align}\label{expansion}
&\log\left[\frac{\sin(\frac{x}{1+\delta x/\pi})}{\sin x}\,\frac{\sin(\frac{x}{1-\delta x/\pi})}{\sin x}\right]
=(2x \cot x-\tfrac{x^2}{\sin^2x})\,\tfrac{\delta x^2}{\pi^2}\,+\,\mathcal{O}(\delta x^4)\,.
\end{align}
With \eqref{substitutions} and \eqref{expansion} we can
express the leading contribution 
to $\log {\cal P}_2$ for large $k$ and $N$ as
\begin{align}\label{sumexpression}
\log{\cal P}_2=&\;\sum_{m=1}^N\big(\nu-\tfrac{x}{\pi}\big)\big(
2x\cot x-\tfrac{x^2}{\sin^2x}\big)\,\tfrac{\delta x}{\pi}
\, +\,\mathcal{O}(\tfrac{\nu^2}{N^2})\,.
\end{align}
The sum is convergent as long as every $x$ is smaller 
than $\pi$,
which means that the expansion is valid in 
the case $0\leq\nu<1$.
In the 't Hooft limit the sum becomes an integral. 
Since the error term is of order $1/N^2$, the sum will 
yield the correction up to first order in $1/N$. 
By the Euler-Maclaurin formula we obtain
\begin{align}\label{logP2leadingorderintegral}
\log{\cal P}_2&=\frac{1}{\pi}\int_0^{\pi\nu}\big(\nu-\tfrac{x}{\pi}\big)\big(
2x\cot x-\tfrac{x^2}{\sin^2x}\big)\,dx\,+\,\frac{\nu^2}{2N}\,+\,\mathcal{O}(N^{-2})\nonumber\\
&=\frac{1}{\pi^2}\int_0^{\pi\nu}x^2\cot x\,dx\,+\,\frac{\nu^2}{2N}\,+\,\mathcal{O}(N^{-2})\qquad(\nu<1)\,.
\end{align}
Combining the results \eqref{fulllogP1} 
and \eqref{logP2leadingorderintegral} we find that
\begin{align}\label{gfactorresult}
g^2={\cal P}_1{\cal P}_2&=\exp\left[
\pi\int_0^{\nu}\lambda^2\cot (\pi\lambda)\,d\lambda
\;+\;\frac{\nu^2}{N}\,+\,\mathcal{O}(N^{-2})\right]\,.
\end{align}
In the Hooft limit we therefore have
\begin{equation}
\frac{d}{d\nu}\log g^2=\pi\nu^2\cot(\pi\nu)\,.
\end{equation} 
After including the factor of 2 for the complex field, this
agrees precisely with the bulk result~\eqref{eq:g-d=2}.

\subsubsection{Matching of coefficients for 
two-point functions}
We can also use the recipe of 
section~\ref{sec:GaiottosRGInterface} to check the 
coefficients in the two-point functions of 
section~\ref{sec:2ptfcn}.
The bulk scalar field is dual on the IR side of the interface to 
the CFT operator $\varphi_{+}=\Phi_{(f,0)}^{{IR}}$, and to $\varphi_-=\Phi_{(0,f)}^{UV}$ on the 
UV side, where $f$ denotes the fundamental representation of $\su(N)$. 
The conformal dimensions of $\Phi_{(f,0)}^{{IR}}$ and
$\Phi_{(0,f)}^{{UV}}$ for finite $N$ and $k$ are
\begin{equation}
\Delta_{(f,0)}^{IR}=\frac{N-1}{N}\left(1+\frac{N+1}{N+k}\right)\,,
\qquad \Delta_{(0,f)}^{UV}=\frac{N-1}{N}\left(1-\frac{N+1}{N+k+1}\right)\,.
\end{equation}
The first coefficient we would like to match is 
the constant $B$ in~\eqref{GreensFunctionOneSide}.
Writing the OPE of the scalar field in the IR 
as~$\Phi_{(f,0)}^{{IR}}\times 
\Phi_{(f,0)}^{{IR}}\sim 1+C'_{IR}\Phi_{\rm (adj,0)}^{IR}$,
the constant $B$ is given by the expression
\begin{equation}
B=C'_{IR}\,g^{-1}\langle\Phi_{\rm (adj,0)}^{{IR}}{\rm id}^{UV}|RG\rangle\,.
\end{equation}
The operator $\Phi_{\rm (adj,0)}^{IR}$ 
corresponds to the double trace perturbation 
in the IR.

For finite $N$ and $k$, the value of $C'_{IR}$ 
can be obtained, {\it e.g.}, from Coulomb gas 
methods~\cite{Dotsenko:1984nm,Chang:2011vka}. 
Since this calculation is not in the focus of this paper 
we refrain 
from performing it here, and only point out that in the 't~Hooft 
limit the OPE coefficients of UV and IR coincide, with~$C'$ 
approaching~1. 
The coefficient~$C$ goes to~2, in agreement with 
condition~\eqref{CCprimeCondition}. 

Now consider the overlap of the IR 
operator~$\Phi_{\rm (adj,0)}^{IR}$ with the identity in the UV. 
In the numerator $\su(N)_{k-1}\otimes \su(N)_{1'}$ of the IR 
coset, the chiral state corresponding to this operator 
can be written as
\begin{equation}\label{IRstate}
\ket{\phi^{IR}_{\rm (adj,0)}} = \frac{1}{\sqrt{\cal N}}
\left(J_{-1}^{(k-1)a}-(2N+k-1)\,J_{-1}^{(1')a}\right)
\ket{J^{(k-1)}_a}\,,
\end{equation}
with normalization constant
\begin{equation}
{\cal N}=(N^2-1)(2N+k-1)(2N+k)\,.
\end{equation}
In \eqref{IRstate}, a sum over the 
indices~$a$ of the currents is implied, and indices are 
raised and lowered with the Killing form $K^{ab}$. 
The state $\ket{J^{(k-1)}_a}$ is the corresponding Virasoro
highest-weight state present in the $\su(N)_{k-1}$ 
adjoint representation. 

Following the recipe of section~\ref{GaiottoInterfaceConstruction} 
we compute the overlap
\begin{equation}\label{overlap}
\cor{({\rm id}^{UV}\tilde{\phi}^{IR}_{\rm (adj,0)})\,
\mathbb{Z}_2(\tilde{{\rm id}}^{UV}\phi^{IR}_{\rm (adj,0)})}
\end{equation}
by replacing $J^{1}_{-1}$ by $J^{1'}_{-1}$
in the $\mathbb{Z}_2$-flipped state. This yields
\begin{equation}\label{overlapfactor}
\cor{({\rm id}^{UV}\tilde{\phi}^{IR}_{\rm (adj,0)})\,
\mathbb{Z}_2(\phi^{IR}_{\rm (adj,0)}\tilde{{\rm id}}^{UV})}
=\frac{1}{2N+k}\;=\;\frac{1-\nu}{1+\nu}\,\frac{1}{k}
\,.
\end{equation}
The Ishibashi state coefficient is
\begin{equation}\label{Smatrixfactor}
\frac{\sqrt{S_{{\rm adj},0}^{(k-1)}S_{00}^{(k+1)}}}{S_{00}^{(k)}}\,.
\end{equation}
Its computation is similar to that of the $g$ factor in the section
above. Using equation~\req{Smatrixelement} of the appendix we find that
\begin{equation}
\frac
    {S_{{\rm adj},0}^{k-1} S_{0,0}^{k+1}}
    {(S_{0,0}^{k})^2}
=
g^2\times 
\frac
    {\sin(\frac{\pi(N+1)}{k+N-1})\sin(\frac{\pi(N-1)}{k+N-1}))}
    {(\sin(\frac{\pi}{k+N-1}))^2}
\,,
\end{equation}
where we used the expression for the $g$ factor from 
the previous section.
In the 't Hooft limit, 
the right-most factor goes as
\begin{equation}
\frac{\sin(\frac{\pi(N+1)}{k+N-1})\sin(\frac{\pi(N-1)}{k+N-1})}%
{(\sin(\frac{\pi}{k+N-1}))^2}\,=\frac{k^2\sin^2(\pi\nu)}{\pi^2(1-\nu)^2}\,+\,\mathcal{O}(k)\,.
\end{equation}
The two factors \eqref{overlap} and \eqref{Smatrixfactor} therefore combine into
\begin{align}
\cor{\Phi^{IR}_{\rm (adj,0)}{\rm id}^{UV}|RG} &= g\, \frac{\sqrt{\sin(\frac{\pi(N+1)}{k+N-1})\sin(\frac{\pi(N-1)}{k+N-1})}}%
{\sin(\frac{\pi}{k+N-1})}\,\frac{1}{2N+k} \nonumber\\
&=\frac{g}{\pi}\sin(\pi\nu)\frac{1}{1+\nu}\,+\,\mathcal{O}(k^{-1},N^{-1})\,,
\end{align}
and we therefore have
\begin{equation}
B=\frac{\sin(\pi\nu)}{\pi(1+\nu)}\,.
\end{equation}
Comparing with~\eqref{eq:B-coefficient} we observe that we have a precise match.

Choosing the two insertions to be on the UV side can be done in the analogous way,
and in the limit merely results in the replacement $\nu\mapsto-\nu$.

It is also straightforward to verify the 
overlap of the scalar across the interface we found 
in section~\ref{sec:pmoverlap}. In the UV theory,
the chiral part of the scalar $\varphi_-$, corresponding 
to $\Phi_{(0,f)}^{UV}$, can be written as a state in the 
numerator of the UV coset as
\begin{equation}\label{fundamentalUVstate}
\ket{\phi^{UV}_{(0,f)}}=\ket{\omega_1}^{(1)}\,,
\end{equation}
where $\omega_1$ denotes the first fundamental weight (which
is the highest weight in the fundamental representation) 
of $\su(N)$, and $\ket{\omega_1}^{(1)}$ is the highest weight state
of the fundamental representation of $\su(N)_1$. In order to have a
non-vanishing overlap we insert the conjugate of the scalar 
$\varphi_+$ in the IR, corresponding to $\Phi_{(\bar{f},0)}^{IR}$.
The chiral state in the IR coset numerator lies
in the product $\bar{f}^{(k-1)}\otimes f^{(1')}$ of the 
antifundamental representation of $su(N_{k-1})$ and the
fundamental representation of $su(N_{1'})$. 
It is given by
\begin{equation}\label{fundamentalIRstate}
\ket{\phi^{IR}_{(\bar{f},0)}}=\frac{1}{\sqrt{N}}\sum_{i=1}^N(-1)^i\ket{\omega_{N-1}-\alpha_{N-1}-\ldots-\alpha_i}^{(k-1)}\ket{\omega_1-\alpha_1-\ldots-\alpha_{i-1}}^{(1')}\,,
\end{equation}  
where $\omega_{N-1}$ is the highest weight of 
the antifundamental representation, 
$\alpha_i$ are the simple roots of $\su(N)$,
and $\ket{\lambda}^{(k)}$ denotes the basis state of weight $\lambda$
in the fundamental (or anti-fundamental) representation of $\su(N)_k$.
For finite $N$ and $k$, the overlap coefficient of one scaler field insertion
on each side of the interface is
\begin{equation}\label{pmoverlap}
\langle \Phi_{(0,f)}^{UV}\Phi_{(f,0)}^{IR}|RG\rangle = 
\frac{\sqrt{S^{(k-1)}_{0f}S^{(k+1)}_{0f}}}{S^{(k)}_{00}}
\langle(\phi_{(0,f)}^{UV}\tilde{\phi}_{(\bar{f},0)}^{IR})%
\mathbb{Z}_2(\tilde{\phi}_{(0,f)}^{UV}\phi_{(\bar{f},0)}^{IR})\rangle\,.
\end{equation}
In the prefactor of modular $S$ matrices we notice that
for any level $k$,
\begin{align}
\frac{S^{(k)}_{0f}}{S^{(k)}_{00}}&=\prod_{m=1}^{N-1}\frac{\sin\frac{\pi(m+1)}{N+k}}{\sin\frac{\pi m}{N+k}}
=\frac{\sin(\pi\nu)}{\sin(\pi\nu/N)}
=\left(\frac{\sin(\pi\nu)}{\pi\nu}\,N+\mathcal{O}(N^{-1})\right)\,.
\end{align}
Using the explicit expressions \eqref{fundamentalUVstate} and \eqref{fundamentalIRstate}, the
other factor in \eqref{pmoverlap} becomes 
\begin{align}
&\langle(\phi_{(0,f)}^{UV}\tilde{\phi}_{(\bar{f},0)}^{IR})%
\mathbb{Z}_2(\tilde{\phi}_{(0,f)}^{UV}\phi_{(\bar{f},0)}^{IR})\rangle =\frac{1}{N}\,.
\end{align}
The RG overlap in the 't Hooft limit is therefore
\begin{equation}
\langle \Phi_{(0,f)}^{UV}\Phi_{(f,0)}^{IR}|RG\rangle =g\,\frac{\sin(\pi\nu)}{\pi\nu}\,.
\end{equation}
Dividing by $g=\corr{1}$ and including a factor of 2 for the complex scalar, this is indeed what we obtain as coefficient from~\eqref{PMoverlapCanonicalNorm}.

\section{Conclusions and Discussion}
In this paper, we gave a semi-classical holographic construction of
double trace interfaces -- RG interfaces associated to an RG flow initiated by double trace deformation. 
We discussed methods for constructing double trace interfaces of 
any shape and computing observables using mixed boundary value problem techniques.
We gave a simple integral representation for the bulk Green's 
function associated to a spherical interface, as well as the 
bulk-boundary propagators and CFT two-point correlation functions in closed form.
From these results we obtained the leading contribution of the 
spherical defect to the CFT partition function (yielding for $d=2$ the boundary entropy). 

Double trace interfaces have arisen previously
in concrete systems of interest, allowing us to test our 
gravitational results against CFT computations. 
We derived the two-point function in the presence of double trace 
interfaces in conformal perturbation theory, and showed that the 
result matches the weak-coupling limit of our gravitational 
computation in the large-$N$ limit, where the single trace operator becomes a generalized free field.
This result generalizes the special case of a Wilson-Fisher/free field interface near $d=4$, studied in~\cite{Gliozzi:2015qsa} using bootstrap methods.
It would be interesting to compute the correlator at large $N$ in the most physically relevant dimension $d=3$.
This should be doable by standard methods; we leave this to future work.

In $d=2$, the $\cW_N$ minimal model RG defects constructed 
in~\cite{Gaiotto:2012np} are realized as double trace interfaces 
within the higher spin gravity/WCFT proposal of~\cite{Gaberdiel:2010pz}.
Using our results, we were able to compute several interface 
overlap coefficients in the semi-classical limit. 
We computed the same coefficients using the exact results 
of~\cite{Gaiotto:2012np} and showed that they coincide at large $N$.
Furthermore, we computed the exact boundary $g$-factor in these 
models, and showed that its large $N$ limit is reproduced by our one-loop gravitational result.

\subsection*{Questions and future directions}
There are several further observables associated to double trace interfaces that would be interesting to compute.
One is the 
leading (one-loop) correction to the stress tensor two-point function 
(which in $d=2$ reduces to the transmission/reflection coefficients of~\cite{Quella:2006de}) and other operators,
and the leading (classical) contribution to the higher-point functions.

A further question, 
of interest for the theory of conformal interfaces, would be to study the fusion of double trace interfaces.
This computation was outlined in section~\ref{sec:defect fusion}.
 
There are two further general points of possible interest we would like to mention.
The first is related to defect conformal bootstrap.
For free fields, the work of~\cite{Liendo:2012hy} showed that the two-point function for Dirichlet and Neumann boundary conditions of a free field could be reproduced by imposing crossing symmetry.
We constructed the large-$N$ spectrum of non-trivial defect operators, and saw that the conformal block decomposition of our two-point functions closes on these operators in the boundary channel, and on double trace operators in the bulk channel.
It is interesting to ask whether our two-point functions are the \emph{unique} solution to the crossing equations that can be generated in this way at large $N$; it is further possible that, using this boundary spectrum as a starting point, one could push the analytic bootstrap results of~\cite{Gliozzi:2015qsa} past leading order in $\epsilon$. 
It is also tempting to apply Mellin bootstrap~\cite{Gopakumar:2016cpb} methods to this problem, since there the effects of double trace operators are included automatically in the Mellin space representation.

The second point is that the match between the gravitational partition function and the $g$-factor for Gaiotto's defect provides further evidence for the proposal of~\cite{Gaberdiel:2010pz} that Zamolodchikov's integrable RG flow is implemented holographically as a double trace deformation.
The starting-point of this RG flow is described on the one hand by the alternative quantization, but according to the original duality it should be described also by a higher spin gravity with the standard quantization but (at finite $N$) a slightly different value of $\nu$. 
This suggests a duality between distinct higher spin gravity theories.
It was shown in~\cite{Giombi:2013fka} that the one-loop correction to the central charge is also consistent with this hypothesis.
It would be of interest to pursue this question further.

\section*{Acknowledgments}
We thank Shouvik Datta for collaboration on early stages of this work. 
CSC would like to thank Ilka Brunner and Enrico Brehm for discussions.
CMT wishes to acknowledge the support of the Thousand Young Talents Program and Fudan University.

\appendix

\section{Coordinate systems on hyperbolic space}
\label{sec:coordinate systems}
We word exclusively in Euclidean signature in this paper, so we are concerned with hyperbolic space $H^{d+1}$ in $d+1$ dimensions.

\subsection{Standard coordinates}
The standard Poincar\'e patch metric for hyperbolic space of unit radius is
\be
ds^2_{H^{d+1}} = \frac{du^2 + d\chi^i d\chi^i}{u^2} \,.
\ee
It is also natural for us to work with spherical defects, in which case 
it is helpful to use radial coordinates on the flat boundary:
\be
ds^2_{H^{d+1}} = \frac{du^2 + dr^2 + r^2 d\Omega_{d-1}^2}{u^2}\,.
\label{radial g}
\ee
The boundary of $H^{d+1}$ is located at $u=0$.

We also work with the Poincar\'e ball model, whose metric is
\be
ds^2 = d\rho^2 + \sinh^2\!\rho\,ds^2_{S^d} \,.
\ee
The defining function is $2 e^{-\rho}$, and the conformal boundary is the sphere located at $\rho\to\infty$.

\subsection{Janus Coordinates}
The Janus decomposition is a slicing of $H^{d+1}$ by surfaces whose geometry
is $H^{d}$.
The standard form of the Janus decomposition is
\be
ds^2_{H^{d+1}} = \frac{1}{\cos^2\mu} (d\mu^2 + ds^2_{H^d}) \,.
\ee
In this coordinate system, the boundary is split into two components, 
located at $\mu = \pm\frac{\pi}{2}$.
The interface between the two boundary components is located at the boundary
of $H^d$.

We work rather in terms of the following coordinate systems:
\be
\sinh\beta=\tan\mu \,,
\qquad\qquad
z = \frac{1}{2}(1-\sin\mu)
\ee
in terms of which
\begin{align}
ds_{H^{d+1}} 
&= \frac{1}{4z^2(1-z)^2}(dz^2 + z(1-z)\,ds^2_{H^d})
\label{z metric}\\
&= d\beta^2 + \cosh^2\!\beta\;ds_{H^d}^2 \,.
\label{kappa metric}
\end{align}

\noindent
{\bf Defining function.}\;
Choosing a member of the conformal class of metrics on the boundary is equivalent to choosing a defining functional $\zeta$ with the following properties.
\begin{enumerate}
\item In a neighborhood $U$ of the boundary, $\zeta$ vanishes on the boundary but nowhere else.
\item In $U$, the metric can be written in Fefferman-Graham form
\be
ds^2 = \frac{d\zeta^2+\gamma_{ij}(\zeta,x)dx^i dx^j}{\zeta^2} \,,
\ee
where $(\zeta,x^i)$ form a coordinate system, and $\lim_{\zeta\to 0}\gamma_{ij}(\zeta,x)$ yields a non-degenerate metric for all $x$.
\end{enumerate}
The defining function adapted to Janus coordinates is $\zeta = \sqrt{4z(1-z)}$, inducing the boundary metric
$ds^2_{\p M} = ds^2_{H^d}$.
Note that $\zeta$ is not a coordinate at $z=\tfrac{1}{2}$ since it is invariant under $z\mapsto 1-z$, but it is a good coordinate in a neighborhood of either boundary component.

\subsection{\texorpdfstring%
    {Coordinates on $H^d$}
    {Coordinates on Hd}
}
We use the following coordinate systems for the $H^d$ slices of the Janus
geometry.
The standard Poincar\'e patch and ball metrics are
\be
ds_{H^d}^2 
= \frac{dy^2 + d\vec x^2}{y^2} 
= \frac{4(d\hat r^2+\hat r^2d\Omega_{d-1}^2)}{(1-\hat r^2)^2} \,.
\label{eq:Hd pp}
\ee
Rather than $\hat r$, we find the following coordinate more useful:
\be
w = \frac{\hat r^2}{1-\hat r^2} \,,
\ee
in terms of which
\begin{align}
ds^2_{H^d} &= \frac{dw^2}{w(1+w)} + 4w(1+w)d\Omega_{d-1}^2 
\label{slice-metric}
\,.
\end{align}

\section{Bulk-boundary propagator from dual integral equations}
\label{sec:K as mbvp}
The purpose of this section is to explicitly derive an expression for the mixed bulk-boundary correlator by solving a mixed boundary value problem.
The setup is as follows: the interface is the sphere $|\chi|=R$, which separates the $\cft_+$ region in the interior from the $\cft_-$ region of the exterior.
Throughout this section we use the Poinar\'e patch coordinates $(u,\chi)$.

Here we compute the mixed bulk-boundary propagator $K_+(u,\chi;\chi')$ for $|\chi'|<R$.
We will do so by imposing properties \ref{K condition 1}-\ref{K condition 3} in order.
Property~\ref{K condition 1} can be satisfied by expanding $K_+$ in solutions to the wave equation with an unspecified coefficient function $\psi$,
\begin{align}
K_+(u,\chi;\chi') &= \sum_{\ell} Y_{\ell}(\hat \chi,\hat \chi')
\cK_\ell(u,r;r') \\
\cK_\ell(u,r;r') &= 
\frac{1}{r^{d/2-1}}
\int_0^\infty d\xi\, 
    \psi_\ell(\xi) \,
    u^{d/2}K_\nu(\xi u) 
    J_{m_\ell}(\xi r)
\,.
\label{K-integral}
\end{align}
Here $m_\ell = \ell+\tfrac{d}{2}-1$, $r=|\chi|$, and $\hat \chi = \frac{\chi}{r}$ is the unit vector in the $\chi$ direction.
This expansion is in terms of boundary spherical waves $r^{1-d/2}Y_{\ell}(\hat \chi,\hat \chi')J_{m_\ell}(\xi r)$ of Laplacian eigenvalue $\xi^2$, in which $Y_{\ell}(\hat \chi,\hat \chi')$ denotes the spherical harmonic rotationally symmetric around the $\hat \chi'$ axis with Laplacian eigenvalue $-\ell(\ell+d-2)$ ($\ell=0,1,\ldots$).
Using
\be
[u^{d/2}K_\nu(\xi u)]_{\Delta_-} = a_0\xi^{-\nu}
\qquad
[u^{d/2}K_\nu(\xi u)]_{\Delta_+} = b_0\xi^{\nu}
\qquad
a_0 = 2^{\nu-1}\Gamma(\nu)
\qquad
b_0 = -\frac{\Gamma(1-\nu)}{\nu 2^{\nu+1}}
\ee
we obtain the asymptotics of $\cK$ as $\ppu\to 0$,
\begin{align}
[\cK_\ell]_{\Delta_-} &= \frac{a_0}{r^{d/2-1}}\int_0^\infty d\xi\,
\xi^{-\nu} J_{m_\ell}(\xi r)\psi_{m_\ell}(\xi) \\
[\cK_\ell]_{\Delta_+} &= \frac{b_0}{r^{d/2-1}}\int_0^\infty d\xi\,
\xi^{\nu} J_{m_\ell}(\xi r)\psi_{m_\ell}(\xi)\,.
\end{align}

Property \ref{K condition 2} says that $[\cK]_{\Delta_+}$ should vanish when $r>R$. 
This can be guaranteed by imposing the ansatz
\be
\psi_\ell(\xi) = \xi\int_0^R ds\, g_\ell(s) J_{m_\ell-\nu}(\xi s) \,,
\label{eq:psi to g}
\ee
a fact verified as follows.
Since $\nu<1$, this integral exists provided $g_m(s)$ is bounded on $[0,R]$.
Using the relation 
\be
\frac{d}{ds}\left[ s^{\lambda}J_\lambda(\xi s) \right]
= \xi s^\lambda J_{\lambda-1}(\xi s)
\ee
we see that
\begin{align}
[\cK_\ell]_{\Delta_+} &= \frac{b_0}{r^{d/2-1}} 
\int_0^R \frac{ds}{s^{m_\ell+1-\nu}}g_\ell(s) 
\frac{d}{ds}\left[ s^{m_\ell-\nu+1}
\int_0^\infty d\xi\, \xi^\nu J_{m_\ell}(\xi r)
J_{m_\ell-\nu+1}(\xi s) \right] \,.
\end{align}
The inner integral is a Weber-Schafheitlin discontinuous integral (see \textit{e.g.} (11.4.33) of \cite{Abramowitz}), and takes the value
\be
\int_0^\infty d\xi\, \xi^\nu J_{m_\ell}(\xi r)
J_{m_\ell-\nu+1}(\xi s)
= \frac{2^\nu}{\Gamma(1-\nu)}\frac{s^{\nu-m_\ell-1}r^{m_\ell}}{(s^2-r^2)^\nu} 
\theta(s-r) \,.
\label{wsdi}
\ee
We thus find
\be
[\cK_\ell]_{\Delta_+} = 
\frac{2^\nu b_0}{\Gamma(1-\nu)}
r^\ell
\int_0^R \frac{ds}{s^{\ell+d/2-\nu}} g_\ell(s)
\frac{d}{ds}\left[ \frac{\theta(s-r)}{(s^2-r^2)^\nu} \right] \,.
\label{K-delta+}
\ee
In particular, $[\cK_\ell]_{\Delta_+}$ vanishes for $r>R$, and so our ansatz guarantees that \ref{K condition 2} is satisfied.

We must also impose the condition 
$[\cK_\ell]_{\Delta_-} = \delta^{(d)}(x-x')$ for $r<R$.
Inserting our ansatz gives
\be
[\cK_\ell]_{\Delta_-} = \frac{a_0}{r^{d/2-1}} \int_0^R ds\, g_\ell(s)
\int_0^\infty d\xi\, \xi^{1-\nu}J_{m_\ell}(\xi r)J_{m_\ell-\nu}(\xi s) \,.
\label{eq:Kdel-}
\ee
Once again the inner integral is a Weber-Schafheitlin discontinous integral, and takes the value
\be
\int_0^\infty d\xi\, \xi^{1-\nu}J_{m_\ell}(\xi r)J_{m_\ell-\nu}(\xi s)
= \frac{1}{a_0}\frac{s^{m_\ell-\nu}r^{-m_\ell}}{(r^2-s^2)^{1-\nu}}\theta(r-s)
\label{eq:W-S minus}
\ee
so that for $r<R$,
\be
[\cK_\ell]_{\Delta_-} = 
\frac{1}{r^{\ell+d-2}}\int_0^r ds\, g_\ell(s) 
\frac{s^{m_\ell-\nu}}{(r^2-s^2)^{1-\nu}}\,.
\ee
Let us now impose the condition that $[\cK_\ell]_{\Delta_-} = \phi_\ell(r)$, \textit{i.e.,}
\be
\int_0^r ds\, g_\ell(s) \frac{s^{m_\ell-\nu}}{(r^2-s^2)^{1-\nu}} 
= r^{\ell+d-2}\phi_\ell(r)\,.
\label{eq:g integral equation}
\ee
This is an integral equation of Abel type.
For $0<\alpha<1$, the equation
\be
\int_0^r \frac{f(t)dt}{[h(r)-h(t)]^\alpha} = g(r)
\ee
has solution 
\be
f(s) 
= \frac{\sin(\pi\alpha)}{\pi}\frac{d}{ds}\int_0^s 
\frac{h'(t)g(t)du}{[h(s)-h(t)]^{1-\alpha}} 
\ee
(see \textit{e.g.} (2.3.2) of \cite{Sneddon}).
With the substitutions
\be
\alpha \mapsto 1-\nu
\qquad
h(t) \mapsto t^2
\qquad
f(s)\mapsto s^{m_\ell-\nu}g_\ell(s)
\qquad
g(r)\mapsto r^{\ell+d-2}\phi_\ell(r)
\ee
we obtain
\be
    g_\ell(s) = \frac{2}{s^{m_\ell-\nu}}\frac{\sin(\pi\nu)}{\pi}
        \frac{d}{ds}\int_0^s \frac{u^{\ell+d-1}\phi_\ell(u) du}{(s^2-u^2)^\nu} \,.
\label{eq:g solution}
\ee

Our particular condition is $[\cK]_{\Delta_-} = \delta^{(d)}(x-x')$.
The delta function can be expanded
\be
\delta^{(d)}(x-x') = \frac{\delta(r-r')}{r^{d-2}}
\sum_\ell c_\ell Y_\ell(\hat x,\hat x')
\ee
where 
\be
c_\ell=\frac{1}{\cN_\ell}\overline{Y_\ell(\hat x',\hat x')}
\qquad{\rm with}\qquad
\cN_\ell=\int_{S^{d-1}} d\Omega_{d-1}(\hat x)|Y_\ell(\hat x)|^2 \,.
\ee
This means that $\phi_\ell(r) = \frac{c_\ell}{r^{d-1}}\delta(r-r')$, and therefore
\be
g_\ell(s) = 2\,c_\ell\frac{\sin(\pi\nu)}{\pi} r'^\ell s^{\nu-\ell-d/2+1} 
\frac{d}{ds} \left[ \frac{\theta(s-r')}{(s^2-r'^2)^\nu} \right] \,.
\label{g-function}
\ee
Together with equations \req{K-integral} and \req{eq:psi to g}, this yields an explicit integral representation for the mixed bulk-boundary propagator $K_+$.
The relevant integrals can be evaluated in terms of hypergeometric functions, and one can verify that (up to a change in conformal frame) the result matches~\req{eq:K+-}.

\section{\texorpdfstring%
    {Spectral decomposition on $H^d$}
    {Spectral decomposition on Hd}
}
\label{sec:spectral decomposition}
Start with the metric \req{slice-metric} on $H^d$,
\be
ds^2 = \frac{dw^2}{w(1+w)} + 4w(1+w)d\Omega^2_{d-1}(\gamma) \,,
\qquad 
w\in[0,\infty) 
\ee
with $\gamma$ coordinates on $S^{d-1}$.
We look for solutions to the equation
\be
-\nabla_{H^d}^2\Psi = \lambda\Psi \,.
\ee
Let $\ell$ index the harmonics $Y_\ell$ on the unit sphere $S^{d-1}$, and 
denote by $L_\ell$ the eigenvalues of $-\nabla^2_{S^{d-1}}$;
$L_\ell = k(k+d-2)$ for an integer $k$.
Decomposing $\Psi=Y_\ell\psi(w)$, we have
\be
\bigr([w(1+w)]^{1-d/2}\frac{d}{dw}[w(1+w)]^{d/2}\frac{d}{dw}
-[4w(1+w)]^{-1}L_\ell + \lambda\bigr)\psi(w) = 0 \,.
\ee
This equation is hypergeometric, and has a unique solution that is finite
as $w\to 0$:
\be
\psi_{k,\sigma}(w) = \cN_{k,\sigma}\,
w^{k/2}(1+w)^{1-(d+k)/2}\,
\hg{\frac{1}{2}+i\sigma}{\frac{1}{2}-i\sigma}{\frac{d}{2}+k}{-w}
\ee
where we have expressed the eigenvalue in the form 
$\lambda=\sigma^2 + \frac{(d-1)^2}{4}$.

We can find the normalized eigenfunctions in the following way.
First of all, let $Y_\ell$ be normalized,
\be
\int_{S^{d-1}}d\Omega_{d-1}(\gamma)\, Y_{\ell}(\gamma)\overline{Y_{\ell'}(\gamma)}
= \delta_{\ell,\ell'} \,.
\ee
The Olevskii transform gives a resolution of the radial delta function
of the form
\begin{multline}
\left(\frac{1+w}{w}\right)^{\frac{d}{2}+k-1}
\delta(w-w')
= \\
\frac{1}{\pi}\int_0^\infty d\sigma\left| 
\frac{\Gamma(\frac{1}{2}+i\sigma)\Gamma(\frac{d-1}{2} + k + i\sigma)}
{\Gamma(2i\sigma)\Gamma(\frac{d}{2}+k)}
\right|^2 
\hg{\frac{1}{2}+i\sigma}{\frac{1}{2}-i\sigma}{\frac{d}{2}+k}{-w}
\hg{\frac{1}{2}+i\sigma}{\frac{1}{2}-i\sigma}{\frac{d}{2}+k}{-w'} 
\end{multline}
Setting
\be
\cN_{k,\sigma}=\frac{2^{(1-d)/2}}{\sqrt\pi}
\frac{\Gamma(\frac{1}{2}+i\sigma)\Gamma(\frac{d-1}{2} + k + i\sigma)}
{\Gamma(2i\sigma)\Gamma(\frac{d}{2}+k)}
\ee
and 
$\Psi_{\ell,\sigma}(w,\gamma)=Y_{\ell}(\gamma)\psi_{k,\sigma}(w)$, 
we obtain the identity
\be
\int_0^\infty d\sigma
\sum_\ell
\overline{\Psi_{\ell,\sigma}(w',\gamma')}
\Psi_{\ell,\sigma}(w,\gamma)
= \frac{1}{\sqrt{g_{H^d}}}\delta(w-w')\delta(\gamma-\gamma')
= \frac{\sqrt{g_{S^{d-1}}}}{\sqrt{g_{H^{d}}}}\delta(w-w')
	\delta(\gamma,\gamma') \,.
\label{delta resolution}
\ee
Similarly, from the inverse Olevskii transform we find that
\be
\int_{H^d}d^dx\sqrt{g_{H^d}}\, 
\Psi_{\ell,\sigma}(x)
\overline{\Psi_{\ell',\sigma'}(x)}
= \delta_{\ell,\ell'}\,\delta(\sigma-\sigma') 
\ee
where $x^i$ denote the coordinates on $H^d$.

Thus the functions $\Psi_{\ell,\sigma}$ form a complete basis for the 
normalizable functions on $H^d$.

\subsubsection*{$\SO(d,1)$-invariant bifunctions}
Our primary interest is in bifunctions on $H^d$, \emph{i.e.} functions
$u(x,x')$ of two points $x,x'\in H^d$ that are symmetric and invariant 
under $\SO(d,1)$, that are also eigenfunctions of the Laplacian.
As with any function, it is possible to expand it with respect to 
eigenfunctions of the Laplacian $u_\lambda(x,x')$ satisfying the same
properties.
At fixed eigenvalue $\lambda_\sigma$, such functions can be decomposed as a
sum over spherical harmonics of the $\Psi_{\ell,\sigma}$ functions:
\be
u_{\lambda_\sigma}(x,x') = 
\sum_{\ell}c_{\ell}(x')
\Psi_{\ell,\sigma}(x) \,.
\ee
Such a function depends only on the hyperbolic distance, and so it suffices
to set $x'=0$ (\emph{i.e.} $w'=0$).
The expression is further rotationally invariant around $x'=0$, which implies
that only the $\ell=0$ mode contributes.
We thus find
\be
u(x,0) = c'\, \Psi_{0,\sigma}(w)  \,;
\ee
of course, for $\ell=0$ there is no dependence on the angular variables
$\gamma$.
To recover the general expression, we simply express the result in terms of
the hyperbolic distance.

The invariant distance between the point $x'$ with $w'=0$ and 
the point $x=(w,\gamma)$ is
\be
d(x,x') = \int_0^w\, [s(1+s)]^{-1/2}ds = 2\sinh^{-1}(w^{1/2}) 
\ee
which gives $4w(1+w) = \sinh^2 d(x,x')$.
A simple expression for the hyperbolic distance in Poincar\'e patch 
coordinates $\frac{dy^2+d\vec x^2}{y^2}$ can be given in terms of the
cross-ratio $\chi_d^2$:
\be
d(x,x') = \cosh^{-1}(1+2\chi_d^2) \,,
\qquad
\chi_d^2 = \frac{(\vec x-\vec x')^2 + (y-y')^2}{4yy'} \,.
\ee
For $w'=0$, $\chi_d^2=w$, so $u(x,0)$ above can be covariantized to 
general $x'$ by replacing $w\mapsto\chi_d^2$.

A bifunction of particular interest for us is
\be
J_\sigma(x,x')=\sum_\ell \Psi_{\ell,\sigma}(x)\overline{\Psi_{\ell,\sigma}(x')} \,.
\ee
Equation \req{delta resolution} implies that 
$\int d\sigma\, J_\sigma(x,x') = \delta(x,x')$, which is invariant under
$\SO(d,1)$ transformations; because $\SO(d,1)$ doesn't mix eigenvalues of
the Laplacian, this implies that $J_\sigma(x,x')$ itself is invariant
under $\SO(d,1)$ transformations.

By acting with a conformal transformation, we can set $x'=0$,
in which case all modes but $k=0$ drop out.
With $Y_0(\gamma)=(\mathrm{vol}\,S^{d-1})^{-1/2}$,
\be
J_\sigma(x,0)=
\frac{1}{\mathrm{vol}\,S^{d-1}}
\frac{2}{2^d\pi}(1+w)^{1-d/2}
\left|\frac{\Gamma(\frac{1}{2}+i\sigma)\Gamma(\frac{d-1}{2} + i\sigma)}
{\Gamma(2i\sigma)\Gamma(\frac{d}{2})}
\right|^2 
\hg{\frac{1}{2}+i\sigma}{\frac{1}{2}-i\sigma}{d/2}{-w} \,.
\ee
As we saw above, we can find its value at general $x'$ by replacing
$w\mapsto \chi_d^2$:
\begin{align}
J_\sigma(x,x')&=
\frac{1}{(4\pi)^{d/2}}\frac{\Gamma(d/2)}{\pi}
\left|\frac{\Gamma(\frac{1}{2}+i\sigma)\Gamma(\frac{d-1}{2} + i\sigma)}
{\Gamma(2i\sigma)\Gamma(\frac{d}{2})}
\right|^2 
(1+\chi_d^2)^{1-d/2}
\hg{\frac{1}{2}+i\sigma}{\frac{1}{2}-i\sigma}{d/2}{-\chi_d^2} 
\label{J} \\
&= 
\frac{1}{(4\pi)^{d/2}}\frac{\Gamma(d/2)}{\pi}
\left|\frac{\Gamma(\frac{1}{2}+i\sigma)\Gamma(\frac{d-1}{2} + i\sigma)}
{\Gamma(2i\sigma)\Gamma(\frac{d}{2})}
\right|^2 
\hg{\frac{d-1}{2}+i\sigma}{\frac{d-1}{2}-i\sigma}{d/2}{-\chi_d^2} \,.
\end{align}
We often require the value at coincidence:
\be
\cN_\sigma := J_\sigma(x,x) = 
\frac{1}{(4\pi)^{d/2}}\frac{\Gamma(d/2)}{\pi}
\left|\frac{\Gamma(\frac{1}{2}+i\sigma)\Gamma(\frac{d-1}{2} + i\sigma)}
{\Gamma(2i\sigma)\Gamma(\frac{d}{2})}
\right|^2 \,.
\ee

\section{Integral transforms}
In Janus coordinates, we make extensive use of a hypergeometric index 
integral transform.
The transform in question is a generalization of the Mehler-Fock transform that was first discovered by Weyl \cite{Weyl}.
His work was largely forgotten, and the same integral transform was later
rediscovered by Titchmarsh \cite{titchmarsh} and Olevskii \cite{olevskii}.

Let $a,c>0$, and $f(x)$ be a sufficiently well-behaved function (say, smooth
and of compact support) on $\R_+$. 
The transform $\cJ_{a,c}\{f\}$ of $f$ is
\be
g(s) = \cJ_{a,c}\{f\}(s) = 
\int_0^\infty dx\, x^{a+c-1}(1+x)^{a-c} \hg{a+is}{a-is}{a+c}{-x}
\, f(x) \,.
\ee
The inversion theorem (see, \emph{e.g.,}~\cite{Neretin}) states that $f(x)$ is recovered by
the following formula:
\be
f(x) = \cJ^{-1}_{a,c}\{g\}(x) = 
\frac{1}{\pi}\int_0^\infty ds\, 
\left|\frac{\Gamma(a+is)\Gamma(c+is)}{\Gamma(2is)\Gamma(a+c)}\right|^2
\hg{a+is}{a-is}{a+c}{-x}\, g(s) \,.
\ee

\section{Hypergeometric functions}
The Gauss hypergeometric function is defined by
\be
\hg abcz = \sum_{n=0}^\infty \frac{(a)_n(b)_n}{(c)_n}\frac{z^n}{n!}
\ee
We also encounter the generalized hypergeometric function
\be
\ghg{p}{q}{a_1,\ldots,a_p}{b_1,\ldots,b_q}{z}
=
\sum_{n=0}^\infty \frac{(a_1)_n\cdots(a_p)_n}{(b_1)_n\cdots(b_q)_n}\frac{z^n}{n!}
\,.
\label{eq:pfq}
\ee

\subsubsection*{Differential equation}
Set $F(z)=\hg abcz$. Then $F$ satisfies
\be
z(1-z)F'' + \bigl[ c - (a+b+1)z \bigr] F' - ab F = 0 \,.
\ee

\subsection{Identities}
%
\subsubsection{Euler identities}
\begin{align}
\hg{a}{b}{c}{z} 
&= (1-z)^{c-a-b}\;\hg{c-a}{c-b}{c}{z} \\
&= (1-z)^{-a}\;\hg{a}{c-b}{c}{-\frac{z}{1-z}} \\
&= (1-z)^{-b}\;\hg{c-a}{b}{c}{-\frac{z}{1-z}}
\end{align}
%

\subsubsection{Kummer's connection formulas}
Defining
\begin{subequations}
\begin{align}
\Phi_1 &= \hg abcz &
\Phi_2 &= z^{1-c}\hg{a-c+1}{b-c+1}{2-c}z \\
\Phi_3 &= \hg ab{a+b-c+1}{1-z} &
\Phi_4 &= (1-z)^{c-a-b}\hg{c-a}{c-b}{c-a-b+1}{1-z} \\
\Phi_5 &= (-z)^{-a}\hg{a}{a-c+1}{a-b+1}{\frac{1}{z}} &
\Phi_6 &= (-z)^{-b}\hg{b-c+1}{b}{b-a+1}{\frac{1}{z}}
\end{align}
\end{subequations}
we have
\begin{subequations}
\begin{align}
\Phi_1 &= \frac{\Gamma(c)\Gamma(c-a-b)}{\Gamma(c-a)\Gamma(c-b)}\Phi_3
    + \frac{\Gamma(c)\Gamma(a+b-c)}{\Gamma(a)\Gamma(b)}\Phi_4 
    \label{eq:hg z->1-z} \\
&= \frac{\Gamma(c)\Gamma(b-a)}{\Gamma(b)\Gamma(c-a)}\Phi_5
    + \frac{\Gamma(c)\Gamma(a-b)}{\Gamma(a)\Gamma(c-b)}\Phi_6 \,.
    \label{eq:hg z->1/z}
\end{align}
\end{subequations}
These are guaranteed to be valid for $0<\Re z<1$ and $\Re z<0$, respectively; for general values of $z$ one needs to take care with the branch cuts.

\subsubsection{Sum relations\label{sec:hg sum}}
The decomposition into conformal blocks of section~\ref{sec:conformal blocks} 
is accomplished using equation (4.3.11) of \cite{erdelyi},
\be
\hg{a}{b}{c}{z}=\sum_{k=0}^\infty \frac{(\alpha)_k(\beta)_k}{k!\,(\gamma+k-1)_k}
(-z)^k {}_4F_3\biggl(\!\begin{array}{c} a, b, \gamma+k-1, -k \\ \alpha, \beta, c \end{array} \,\bigg|\, 1\biggr)
\hg{\alpha+k}{\beta+k}{\gamma+2k}{z}
\,,
\label{eq:hg sum}
\ee
valid for any choice of $\alpha,\beta,\gamma$ such that the identity makes sense.
For our two-point function, ${}_4F_3$ reduces to ${}_3F_2$, and we apply Saalsch\"utz's theorem
\be
{}_3F_2\biggl(\!\begin{array}{c} a, b, -k \\ c, a+b+1-c-k \end{array} \,\bigg|\, 1\biggr)
= \frac{(c-a)_k(c-b)_k}{(c)_k(c-a-b)_k}
\qquad
k = 0,1,2,\ldots
\ee

\subsection{Integrals
\label{sec:integrals}}
When the sum \req{eq:pfq} converges uniformly on $[0,1]$, Taylor expansion together with the beta integral 
implies
\be
\int_0^1 dz\, z^{\mu-1}(1-z)^{\nu-1}
\ghg{p}{q}{a_1,\ldots,a_p}{b_1,\ldots,b_q}{z}
=
\frac{\Gamma(\mu)\Gamma(\nu)}{\Gamma(\mu+\nu)}
\ghg{p+1}{q+1}{\mu,a_1,\ldots,a_p}{\mu+\nu,b_1,\ldots,b_q}{z}
\ee
provided the integral exists.
If the integrand is bounded as $z\to 1$ but diverges at $z=0$, subtraction of the power law divergences is equivalent to performing analytic continuation in $\mu$.
Only a finite number of terms give rise to divergences; the most divergent contribution has the form $-\frac{z^{\mu}}{\mu}$.

\subsection{Index integrals of hypergeometric functions
\label{sec:mellin}}
Let $a_1,a_2,a_3,a_4\in\R$, and consider the integral
\be
\frac{1}{\pi}\int_0^\infty d\sigma
\left|\frac{\Gamma(a_1+i\sigma)\Gamma(a_2+i\sigma)\Gamma(a_3+i\sigma)\Gamma(a_4+i\sigma)}{\Gamma(2i\sigma)}\right|
\ghg{p+2}{q}{a_1+i\sigma,a_1-i\sigma,b_1,\ldots,b_p}{c_1,\ldots,c_q}{z}
\,.
\ee
Expanding in powers of $z$ and using the De Branges-Wilson integral
\be
\frac{1}{\pi}\int_0^\infty d\sigma \left|
\frac{\prod_{i=1}^4\Gamma(a_i+i\sigma)}{\Gamma(2i\sigma)}
\right|^2
= 
\frac{\prod_{i<j}\Gamma(a_i+a_j)}{\Gamma(\sum_i a_i)}
\ee
we see that it equals
\be
\frac{\prod_{i<j}\Gamma(a_i+a_j)}{\Gamma(\sum_i a_i)}
\ghg{p+3}{q+1}{a_1+a_2,a_1+a_3,a_1+a_4,b_1,\ldots,b_p}{\sum_i a_i,c_1,\ldots,c_q}{z}
\,.
\ee

\subsection{Three-term relations for 
\texorpdfstring{${}_3F_2(1)$}{3F2(1)}
\label{sec:3F2 sum rule}}
The purpose of this section is to provide a sum rule for $\f{a_0,a_1,a_2}{b_1,b_2}=\htt{a_0}{a_1}{a_2}{b_1}{b_2}{1}$.
With $\mathbf{a}=\begin{pmatrix}a_0&a_1&a_2\\& b_1&b_2\end{pmatrix}$, define
\be
y(\mathbf{a}) = \frac{\prod_{i=0}^2\sin\pi a_i}{\prod_{i=1}^2\sin\pi b_i}
\f{a_0,a_1,a_2}{b_1,b_2} \,,
\ee
and set
\begin{align}
\tau_1(\mathbf{a}) &= 
\begin{pmatrix}
a_0-b_1+1 & a_1-b_1+1 & a_2-b_1+1 \\
& 2-b_1 & b_2-b_1+1
\end{pmatrix} \\
\tau_2(\mathbf{a}) &= 
\begin{pmatrix}
a_0-b_2+1 & a_1-b_2+1 & a_2-b_2+1 \\
& b_1-b_2+1 & 2-b_2
\end{pmatrix} \,.
\end{align}
We utilize the following standard 3-term relation (see, \emph{e.g.,} \cite{Ebisu}, whose notation we follow):
\be
y(\mathbf{a}) + y(\tau_1(\mathbf{a})) + y(\tau_2(\mathbf{a})) = 0 \,.
\ee
Applying this to
\be
\mathbf{a} = \begin{pmatrix}s\nu&1+s\nu&\Delta_s\\&1+2s\nu&\Delta_s+1+\epsilon\end{pmatrix}
\ee
and taking the limit $\epsilon\to 0$ gives the relation
\be
-\frac{1}{2}\tan(\pi\nu)\sum_{s=\pm 1} s\, \f{s\nu,1+s\nu,\Delta_s}{1+2s\nu,\Delta_s+1}
+ \pi\frac{\Gamma(\Delta_+)\Gamma(\Delta_-)}{\Gamma(\tfrac{d}{2})\Gamma(\tfrac{d}{2}+1)} = 0 \,.
\label{eq:3-term relation}
\ee

\section{\texorpdfstring{Conventions for $\su(N)$}{Conventions for su(N)}}\label{SUNconventions}
Our conventions for $\su(N)$ and its affine algebras 
follow~\cite{DiFrancesco:1997nk}. Here we collect some facts which are important for our section~\ref{sec:d2CFT}.

\medskip
The dimension of $\su(N)$ is $N^2-1$, and the dual Coxeter number is $g^\vee=N$. Bases of generators are denoted $J^a$, $a=1,\ldots,N^2-1$.
The weight and root lattice of $\su(N)$ can be realized in $\mathbb{R}^N$ with standard basis $e_1,\ldots e_N$: The roots are given by $\alpha=e_i-e_j$ for $i\neq j$, and we define the positive roots to be those with $i<j$. A set of simple roots is then provided by $\alpha_i=e_i-e_{i+1}$ for $i=1,\ldots,N-1$. The root lattice consists of all vectors of the form $\sum_{i=1}^N n_ie_i$ with $n_i\in\mathbb{Z}$ and $\sum_in_i=0$. The Weyl vector, given by half the sum of all positive roots, is represented by $\rho=\frac{1}{2}\sum_{i=1}^N(N+1-2i)e_i$. The fundamental weights are
$\omega_i=\sum_{j=1}^ie_j-\frac{i}{N}\sum_{j=1}^Ne_j$ for $i=1,\ldots,N-1$; every weight is given by $\lambda=\sum_{i=1}^{N-1}\lambda_i\omega_i$ with Dynkin labels $\lambda_i$. In our case we need in particular the fundamental and antifundamental representations with highest weights $\omega_1$ and $\omega_{N-1}$, respectively. The fundamental representation contains the weights $\omega_1-\sum_{j=1}^i\alpha_j$, and the antifundamental representation contains the weights $\omega_{N-1}-\sum_{j=1}^i\alpha_{N-j}$ for $i=0,1,\ldots,N-1$ (empty sums are 0). We also need the adjoint representation $\theta$, which has Dynkin labels $1,0,\ldots,0,1$.

\medskip
In the $\su(N)_k$ WZW model, the chiral fields $J^a(z)$ can be decomposed into modes $J^a_n$, $n\in\mathbb{Z}$, where $J^a_0$ act as $-J^a$ on the Virasoro highest weight states $\ket{\lambda}$ labeled by $\su(N)$ weights, and $J^a_n\ket{\lambda}=0$ for $n>0$. The $\ket{\lambda}$ have (chiral) conformal dimension $(\lambda,\lambda+2\rho)/2(k+N)$, where the inner product coincides with the standard one on $\mathbb{R}^N$.
Highest weight operators with respect to the $\su(N)_k$ algebra only occur if $(\lambda,\theta)\leq k$.

The formula for the elements in the first column of the (symmetric) modular $S$ matrix is
\begin{equation}\label{Smatrixelement}
S_{\lambda0}^{(k)}=|\det((\alpha^{\!\vee}_i)_j)|^{-\frac{1}{2}}(k+N)^{-{N-1}{2}}
\prod_{\alpha>0}2\sin\left(\frac{\pi(\alpha,\lambda+\rho)}{k+N}\right)\,.
\end{equation}
In this formula, the $\alpha^{\!\vee}_i$ denote the coroots, which in $\mathbb{R}^N$ coincide with the roots.

\medskip
For the coset $\su(N)_k\otimes \su(N)_1/\su(N)_{k+1}$, we recall that representations are given by a triple $\Lambda=(\lambda^{(k)},\lambda^{(1)},\lambda^{(k+1)})$ of weights of the respective affine $\su(N)$ algebras. The condition that $\lambda^{(k)}+\lambda^{(1)}-\lambda^{(k+1)}$ needs to be in the root lattice then allows precisely one $\lambda^{(1)}$ for a given pair $(\lambda^{(k)},\lambda^{(k+1)})$.



\newpage{\pagestyle{empty}\cleardoublepage}

\end{document}